\documentclass[%
 reprint,
superscriptaddress,
showkeys,
 aps,
 pra,
]{revtex4-1}

\usepackage[margin=1in]{geometry} 
\usepackage{amsmath,amsthm,amssymb}
\usepackage{graphicx}
\usepackage{braket}
\usepackage{dsfont}
\usepackage{comment}
\usepackage{blkarray}
\usepackage{algorithm}
\usepackage[noend]{algpseudocode}
\usepackage{graphicx}
\usepackage{dcolumn}
\usepackage{bm}
\usepackage{amsfonts}
\usepackage{epsf}  
\usepackage{appendix}
\usepackage{mathtools}
\usepackage{bm}

\usepackage{hyperref}
\usepackage[mathlines]{lineno}

\newtheorem{Theorem}{Theorem}
\newtheorem{Definition}{Definition}
\makeatletter
\def\BState{\State\hskip-\ALG@thistlm}

\usepackage{xcolor}

\begin{document}

\preprint{APS/123-QED}

\title{A Quantum Interior-Point Predictor-Corrector Algorithm for Linear Programming}

\author{P. A. M. Casares}
 \email{pabloamo@ucm.es}
 \affiliation{Departamento de F\'isica Te\'orica, Universidad Complutense de Madrid.}
\author{M. A. Martin-Delgado}%
 \email{mardel@ucm.es}
\affiliation{Departamento de F\'isica Te\'orica, Universidad Complutense de Madrid.}%



\date{\today}

\begin{abstract}
We introduce a new quantum optimization algorithm for dense Linear Programming problems, which can be seen as the quantization of the Interior Point Predictor-Corrector algorithm \cite{Predictor-Corrector} using a Quantum Linear System Algorithm \cite{block-encoding_1}. 
The (worst case) work complexity of our method is, up to polylogarithmic factors, $O(L\sqrt{n}(n+m)\overline{||M||_F}\bar{\kappa}\epsilon^{-2})$ for $n$ the number of variables in the cost function, $m$ the number of constraints, $\epsilon^{-1}$ the target precision, $L$ the bit length of the input data, $\overline{||M||_F}$ an upper bound to the Frobenius norm of the linear systems of equations that appear, $||M||_F$, and $\bar{\kappa}$ an upper bound to the condition number $\kappa$ of those systems of equations.
This represents a quantum speed-up in the number $n$ of variables in the cost function  with respect to the comparable classical Interior Point algorithms when the initial matrix of the problem $A$ is dense: if we substitute the quantum part of the algorithm by classical algorithms such as Conjugate Gradient Descent, that would mean the whole algorithm has complexity $O(L\sqrt{n}(n+m)^2\bar{\kappa} \log(\epsilon^{-1}))$, or with exact methods, at least $O(L\sqrt{n}(n+m)^{2.373})$. 
Also, in contrast with any Quantum Linear System Algorithm, the algorithm described in this article outputs a classical description of the solution vector, and the value of the optimal solution.
\end{abstract}

\pacs{Valid PACS appear here}
\maketitle


\section{\label{sec:intro}Introduction}

Linear Programming problems are among the most fundamental optimization problems \cite{Nering-Tucker,Padberg,Murty}.
Applications abound both at personal and professional fronts:
improving a project delivery, scheduling of tasks, analyzing supply chain operations, shelf space optimization, designing better strategies and logistics and scheduling problems in general. Linear Programming is also used in Machine Learning where Supervised Learning works on the basis of linear programming. A system is trained to fit a mathematical model of an objective (cost) function from the labeled input data that later can predict values from unknown test data \cite{Russell-Norvig,MRT}. 
More specifically, linear programming is a method to find the best outcome from a linear function, such as maximum profit or lowest cost, in a mathematical model whose requirements are represented by linear constraints of the variables.
Semi-Definite Programming (SDP) is an extension of Linear Programming where the objective or cost function is formulated with a non-diagonal matrix
and constraints contain more general inequalities \cite{Vandernberghe-Boyd,Todd,Laurent-Rendl,deKlerk}.  

We are in the time of small quantum computers with reduced computational capabilities due to noisy physical qubits \cite{Preskill,Ions,SCQ1,SCQ2}. The challenge of surpassing the power of current and foreseeable classical computers is attracting a lot of attention in the academia \cite{Preskill2,Aaronson-Arkhipov} and in technological companies. This motivates the endeavour of searching for new quantum algorithms beyond the standard ones that spurred the field of quantum computation in the mid 90s (Shor, Grover, etc.) \cite{Shor,Grover,Nielsen-Chuang,GMD}. Only recently, a quantum algorithm for solving SDP problems has been proposed by Brand\~ao and Svore providing us with the first quantum advantage for these optimization problems \cite{Brandao-Svore,QuantumSDP1,QuantumSDP2,QuantumSDP3,QuantumSDP4}.

\subsection{Background on Linear Programming.}
The development of methods to solve Linear Programming problems has a long tradition starting with the Simplex Method \cite{Murty}, which is simple and widely used in practice, but has (in the worst case) exponential time complexity in the number of variables. In 1979 Khachiyan proved that the ellipsoid method ensured (weak) polynomial complexity the number of variables, $O(n^6L)$ \cite{Khachiyan}. However, in practice the ellipsoid algorithm is complicated and not competitive. In 1984 Karamark proposed the first Interior Point algorithm \cite{Karmarkar}, with complexity $O(n^{3.5}L)$. It was more practical than the ellipsoid method and gave rise to a large variety of available Interior Point methods \cite{Potra-Wright}. The best advantage of these methods is that, contrary to what happens in the Simplex Method, Interior Point algorithms have a worst case runtime polynomial in the number of variables. Among them, the Predictor-Corrector Method \cite{Predictor-CorrectorI,Predictor-Corrector} is arguably one of the best procedures to achieve an extremely well-behaved solution, and requires just $O(\sqrt{n}L)$ iterations. However, the Predictor-Corrector method does not explicitly indicate what method should be used to solve the linear system of equations that appear in each iteration, so the solution of the system will depend on what method is used, as can be seen in table \ref{tabla comparativa complejidad}. For further background on Interior Point methods we refer the reader to the review \cite{Potra-Wright}.

\begin{table*}[ht]
\begin{tabular}{|l|l|l|l|}
\hline
Algorithms for Linear Programming &  Work complexity  & Parallelizable? \\ \hline

Pred-Corr. \cite{Predictor-Corrector} + Conjugate Gradient \cite{conjugate_gradient} & $O(L\sqrt{n}(n+m)^2\bar{\kappa} \log(\epsilon^{-1}))$ & $O((n+m)^2)$  \\ \hline
Pred-Corr. \cite{Predictor-Corrector} + Cholesky decomposition \cite{numerical_recipies} & $O(L\sqrt{n}(n+m)^3)$  &   $O((n+m)^2)$  \\ \hline
Pred-Corr. \cite{Predictor-Corrector} + Optimal exact \cite{dense_exact} & $O(L\sqrt{n}(n+m)^{2.373})$  &   $O((n+m)^{2.373})$  \\ \hline
Multiplicative weights \cite{QuantumSDP3} & $O((\sqrt{n}\left(\frac{Rr}{\epsilon}\right)+\sqrt{m}) \left(\frac{Rr}{\epsilon}\right)^{4})$  &  O(1)\\ \hline
Quantum Interior Point with Newton system \cite{kerenidis2018quantum}  &     $O(L\sqrt{n}(n+m)\mu \bar{\kappa}^3 \epsilon^{-2}\log(\epsilon'^{-1}))$  &   $O((n+m)\epsilon^{-2})$  \\ \hline
Pred-Corr. \cite{Predictor-Corrector} + Block-encoding \cite{block-encoding_1} (This algorithm) & $O(L\sqrt{n}(n+m)\overline{||M||_F}\bar{\kappa}\epsilon^{-2})$  &   $O((n+m)\epsilon^{-2})$  \\ \hline
\end{tabular}
\caption{Comparison of complexity of different algorithms\footnote{For completitude we want to mention that recently \cite{nannicini2019fast} has proposed quantum subroutines to speedup the simplex method.} that can be used for solving dense Linear Programming problems; the first three are purely classical whereas the following three are hybrid quantum-classical. It includes only leading-order terms. QLSA stands for a dense Quantum Linear System Algorithm \cite{block-encoding_1}.
Note that the algorithms \cite{QuantumSDP3} and \cite{kerenidis2018quantum} can be applied to more general problems concerning Semidefinite Programming.  
 In the table $\mu \leq \overline{||M||_F} = O(\sqrt{n})$, which reflects the fact that both \cite{kerenidis2018quantum} and our algorithm share common features even if they were developed independently. In reference \cite{QuantumSDP4}, theorem 24, it is proven that for many combinatorial problems the term $(Rr/\epsilon) = O(n+m)$, affecting the complexity of \cite{QuantumSDP3}. With respect to $L$, for many cases the complexity of the Predictor-Corrector method does not depend on $L$ \cite{Average_performance}; but as an upper bound $O(L)$ complexity will be common to all Interior Point methods \cite{Potra-Wright}. In contrast, the Multiplicative weight method will not depend on it.
Finally, the column `Parallelizable?' gives the number of quantum or classical processors that can be used in parallel to solve the problem; the time complexity is divided by the corresponding amount.
}
\label{tabla comparativa complejidad}
\end{table*}

\subsection{Our algorithm}
Here we present a quantum algorithm that relies on the quantization of this method. 
One important feature of our quantum Interior Point algorithm is that it is a hybrid algorithm: partially classical, partially quantum. This feature has become very common and a similar situation occurs with the Brand\~ao-Svore algorithm in SDP, the Quantum Eigen-Solver for quantum chemistry \cite{Aspuru,Mezzacapo,Solano,Innsbruck,ReviewQCh}, and many others, and has the advantage of requiring shorter coherence times. 
The core of the quantization of the Interior Point algorithm relies on the use of the block encoding techniques \cite{block-encoding_1}, that extend the applicability of the Quantum Linear System Algorithm proposed by Harrow, Hassadim and Lloyd (HHL) \cite{HHL} in the case where $A$ is dense, to solve the linear system of equations that appear in the Predictor-Corrector steps.

However, in order to apply the QLSA in the context of Linear Programming, we have to solve several caveats since the straightforward application of it is doomed to failure.

The quantum Interior Point algorithm  we propose benefits from several fundamental properties inherited from the classical Predictor-Corrector algorithm, and has a better performance than other classical Interior Point algorithms. In particular  \cite{Predictor-Corrector}: 
\begin{enumerate}
    \item The Predictor-Corrector method can solve the Linear Programming problem without assuming the existence of feasible or optimal solutions.
    \item If the Linear Programming problem has solution, the loop of this interior point algorithm approaches feasibility and optimality at the same time for both the primal and dual problem, and if the problem is infeasible or unbounded the algorithm detects infeasibility for either the primal or dual problem.
    \item The algorithm can start from any point near the center of the positive orthant.
\end{enumerate}
The notions of feasible, optimal solutions etc. are defined in Sec. \ref{sec:Classical algorithm} where a self-contained review of the Predictor-Corrector method is presented.

The work complexity of the algorithm proposed here is $O(L\sqrt{n}(n+m)\overline{||M||_F}\bar{\kappa}\epsilon^{-2})$, where $n$ is the number of variables of the cost function, $m$ is the number of constraints, $L$ is the bit length of the input data (see Eq. \eqref{encodingL}), $\overline{||M||_F}$ is an upper bound to the Frobenius norm of the linear systems of equations that appear, $\bar{\kappa}$ is an upper bound to the condition numbers of the linear systems of equations that appear in the Predictor-Corrector steps, and $\epsilon^{-1}$ is the precision with which one wants to solve the linear system of equations. To avoid confusion notice that in the text we will call $\epsilon$ the error and $\epsilon^{-1}$ its associated precision, because a low error means high precision and viceversa. The time complexity of the proposed quantum Interior Point algorithm can be reduced from $O(L\sqrt{n}(n+m)\overline{||M||_F}\bar{\kappa}\epsilon^{-2})$ to $O(L\sqrt{n}\overline{||M||_F}\bar{\kappa})$ distributing the work of each iteration between $O((n+m)\epsilon^{-2})$ quantum processors.

If we substituted the QLSA by a classical Linear System Algorithm, the price to pay would be, at least, an $O(\sqrt{n+m})$ increase in the work complexity, as $||M||_F = O(\sqrt{n+m})$ if the spectral norm of $M$ is bounded \cite{DenseHHL}. For example, if we used conjugate gradient descent, the overall algorithm complexity would be $O(L\sqrt{n}(n+m)^2 \bar{\kappa}\log(\epsilon^{-1}))$. Also, if we wanted to use an exact Linear System Algorithm the best we could hope for is the complexity it takes to exactly invert a matrix \cite{dense_exact}, $O((n+m)^{2.373})$ \cite{coppersmith1990matrix}, thus implying an overall work complexity for the algorithm $O(\sqrt{n}(n+m)^{2.373}L)$, that could be parallelized in $(n+m)^{2.373}$ processors to lower the time complexity $O(\sqrt{n}L)$ up to polylogarithmic terms \cite{dense_exact}. A
summary of these results is presented in table \ref{tabla comparativa complejidad}.

It is worth mentioning that our quantization approach to Linear Programming problems is radically different from the method of Brand\~ao and Svore and this comes with several benefits. Namely, the problem of quantising linear programming using multiplicative weight methods \cite{Arora-Kale} as in Brand\~ao-Svore is that they yield an efficiency depending on parameters $R$ and $r$ of the primal and dual problems. In fact, these parameters might depend on the sizes $n,m$ of the cost function, thereby  the real time complexity of the algorithm remains hidden. For instance, for some classes of problems $Rr/\epsilon = O(n)$ according to theorem 24 of \cite{QuantumSDP4}.
Moreover and generically, unless specified, these $R,r$  parameters cannot be computed beforehand, but after running the algorithm (we will have a similar situation with $\bar{\kappa})$. Thus, the real efficiency of the quantum algorithm is masqueraded by overhead factors behaving badly on $R$ and $r$. Their algorithm has nevertheless a good complexity on $n$, $O(\sqrt{n+m})$, but much worse complexity on the precision, $O(\epsilon^{-5})$ for the most recent improvement of the Brand\~ao-Svore algorithm \cite{QuantumSDP3}.


\subsection{\label{sec:structure}Structure of the paper}

The paper has two main sections. The first reviews the Predictor-Corrector algorithm from \cite{Predictor-Corrector}. It is itself divided in subsections where we explain how to initialize and terminate the algorithm, and the main loop.

In the second section we explain the changes we carry out to be able to use the QLSA from \cite{block-encoding_1}. In particular we start with a subsection discussing the condition number and then we focus on how to prepare the initial quantum states for the QLSA and read out the results using a tomography protocol from \cite{kerenidis2017recommendation}. Finally we explain the QLSA, comment on the possibility of quantizing the termination of the algorithm, and devote two subsections to the complexity of the overall algorithm and its comparison with other alternatives, and the possibility of failure.

\begin{figure*}
    \centering
    \includegraphics[width=470pt]{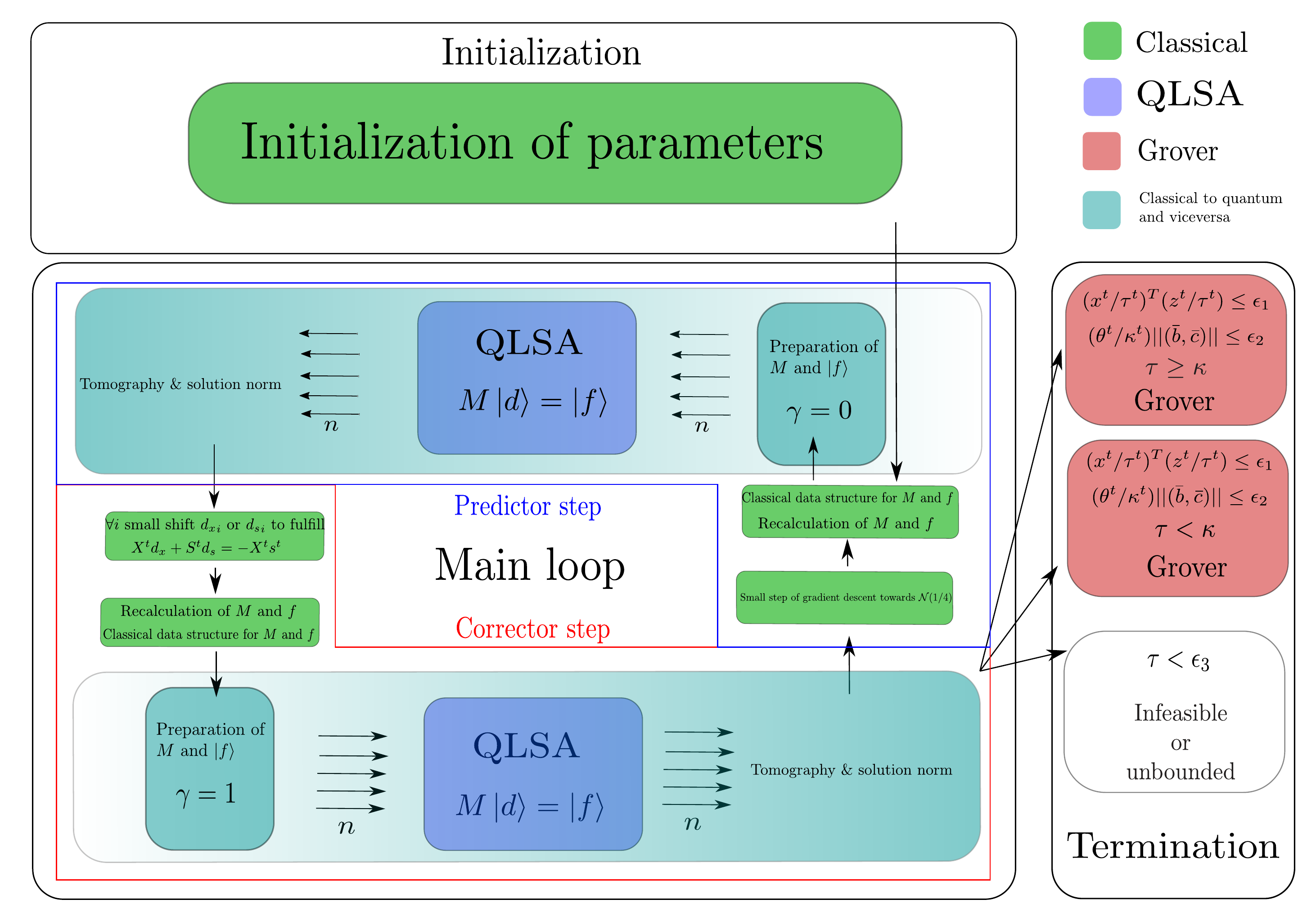}
    \caption{Scheme of the algorithm.}
    \label{fig:scheme}
\end{figure*}

We recommend the reader to first understand the Predictor-Corrector algorithm in its classical form, and then take a look at figures \ref{fig:flow_chart} and \ref{fig:scheme} in order to get an overall impression of the algorithm before trying to understand the technical details.

\section{\label{sec:Classical algorithm}The Predictor-corrector algorithm}

In this section we review the Predictor-Corrector algorithm of Mizuno, Todd and Ye for solving Linear Programming problems \cite{Predictor-Corrector}. As stated in the original article, we will see that it performs $O(\sqrt{n}L)$ iterations of the main loop in the worst case scenario, where $n$ is the number of variables and $L$ the size of the encoding the input data in bits:
\begin{equation}
 L:=\sum_i^n\sum_j^m \lceil \log_2(|a_{ij}|+1)+1\rceil,\label{encodingL}
\end{equation}
for $a_{ij}$ elements of the matrix defining the problem, $A$, that is defined in the following equations \eqref{LP} and \eqref{LD}. Note that $L= O(mn)$. However, in a typical case the number of iterations will not depend on $L$, but will rather be $O(\sqrt{n}\log{n})$ \cite{Average_performance}.

The linear programming problem we want to solve is called primal problem (Linear Problem, LP): Given $A\in \mathbb{R}^{m\times n}$, $\bm{c}\in \mathbb{R}^n$ and $\bm{b}\in \mathbb{R}^m$, find $\bm{x}\in \mathbb{R}^n$ such that:
\begin{subequations}
    \begin{equation}
      \text{minimizes } \bm{c}^T \bm{x}  
    \end{equation}
    \begin{equation}
      \text{subject to } A \bm{x} \geq \bm{b}, \qquad \bm{x}\geq \bm{0}.  
    \end{equation}\label{LP}
\end{subequations}
The dual problem (Dual Problem, DP) has the same solution: finding $y\in\mathbb{R}^m$ such that
\begin{subequations}
    \begin{equation}
        \text{maximizes } \bm{b}^T \bm{y}
    \end{equation}
    \begin{equation}
        \text{subject to } A^T \bm{y} \leq \bm{c}.
    \end{equation}\label{LD}
\end{subequations}
Then, for linear programming problems, the primal-dual gap is $0$:
\begin{equation}
    \bm{b}^T \bm{y} - \bm{c}^T \bm{x} = 0.
    \label{dual gap}
\end{equation}

A usual strategy is to use slack variables to turn all inequality constraints into equality constraints, at the cost of additional constraints. Thus, we can substitute $A^T\bm{y}\leq \bm{c}$ by $A^T\bm{y}+\bm{s}=\bm{c}$, $\bm{s}\geq \bm{0}\in \mathbb{R}^n$ being the slack (dual) variable to the constraint \eqref{LD}. 

\subsection{\label{sec:Initialization}Initialization}

According to the prescription of \cite{Predictor-Corrector}, one way to solve the previous problems \eqref{LP} and \eqref{LD} is to set another problem from which the solution of \eqref{LP} and \eqref{LD} can be easily obtained. This new problem is homogeneous, in the sense that there is a single non-zero constraint, and self-dual, as its dual problem is itself. 
Therefore, let $\bm{x}^0> \bm{0}\in \mathbb{R}^n$, $\bm{s}^0> \bm{0}\in \mathbb{R}^n$, and $\bm{y}^0\in \mathbb{R}^m$ be arbitrary initialization variables which will be chosen later on. Then, formulate the Homogeneous Linear Problem (HLP) as
\begin{equation}
   \min \theta
\end{equation}  
such that ($\bm{x}\geq \bm{0},\tau\geq 0, \tau\in\mathbb{R}$):
\begin{align}
\begin{array}{r@{\,}lr@{\,}lr@{\,}lr@{\,}lcl}
           & &+A&\bm{x} &-\bm{b}&\tau &+\bar{\bm{b}}&\theta & =  & \bm{0}\\
       -A^T&\bm{y}&  &  &+\bm{c}&\tau &-\bar{\bm{c}}&\theta &\geq& \bm{0}\\
       +\bm{b}^T&\bm{y}&-\bm{c}^T&\bm{x}& & & +\bar{z}&\theta &\geq &0\\
        -\bar{\bm{b}}^T& \bm{y}&+\bar{\bm{c}}^T&\bm{x}&-\bar{z}&\tau& & & = & -(\bm{x}^0)^T \bm{s}^0-1
\end{array}
\label{constraints}
\end{align}
with 
\begin{equation}
\begin{split}
    &\bar{\bm{b}}:=\bm{b}-A\bm{x}^0, \qquad \bar{\bm{c}}:=\bm{c}-A^T\bm{y}^0-\bm{s}^0,\\ &\bar{z}:=\bm{c}^T\bm{x}^0+1-\bm{b}^T\bm{y}^0.
    \label{b and c bar}
\end{split}
\end{equation}
The last constraint from \eqref{constraints} is used to impose self-duality. It is also important to remark that $\bar{\bm{b}}$, $\bar{\bm{c}}$ and $\bar{z}$ indicate the infeasibility of the initial primal and dual points, and the dual gap, respectively.

Recall also that we use slack variables to convert inequality constraints into equality constraints. Those slack variables indicate the amount by which the original constraint deviates from an equality.
As we have two inequality constraints, we introduce slack variables $\bm{s}\in \mathbb{R}^n$ for the second constraint in \eqref{constraints} and $k\in \mathbb{R}$ (in \cite{Predictor-Corrector} denoted $\kappa$) for the third:
\begin{align}
  -A^T\bm{y}+\bm{c}\tau-\bar{\bm{c}}\theta -\bm{s}= \bm{0}; \qquad \bm{s}\geq \bm{0} \label{constraint 2 slacked}\\
  \bm{b}^T\bm{y}-\bm{c}^T\bm{x}+\bar{z}\theta-k = 0; \qquad k\geq 0\label{constraint 3 slacked}
\end{align}

This implies that we can rewrite the last constraint in \eqref{constraints} as
\begin{equation}
    (\bm{s}^0)^T\bm{x}+(\bm{x}^0)^T \bm{s}+\tau+k - ((\bm{x}^0)^T \bm{s}^0  +1)\theta= (\bm{x}^0)^T  \bm{s}^0  +1.
\end{equation}
Once we have defined these variables, theorem 2 of \cite{Predictor-Corrector} proves that any point fulfilling 
\begin{equation}
    \bm{y}=\bm{y}^0, \quad \bm{x}= \bm{x}^0>\bm{0},\quad  \bm{s}= \bm{s}^0>\bm{0},\quad  \tau=k=\theta=1.\label{theorem 2 eq 1}
\end{equation}
is a feasible point, and therefore a suitable set of initialization parameters for our algorithm. A particularly simple one can choose is
\begin{equation}
    \bm{y}^0=0_{m\times 1}, \qquad \bm{x}^0=1_{n\times 1}= \bm{s}^0,
\end{equation}
where $1_{n\times 1}=[1,...,1]^T$, and $0_{m\times 1}=[0,...,0]^T$.

\subsection{\label{Main loop}Main loop}
In this section we explain how to set up an iterative method that allows us to get close to the optimal point, following a path along the interior of the feasible region. The original references are \cite{Predictor-CorrectorI,Predictor-Corrector}.
Begin defining $X:=\text{diag}(\bm{x})$ and $S:=\text{diag}(\bm{s})$. Define also $\mathcal{F}_h$ the set of feasible points of (HLP) $\bm{v}=(\bm{y},\bm{x},\tau,\theta, \bm{s},k)$; and $\mathcal{F}_h^0\subset \mathcal{F}_h$ those such that $(\bm{x},\tau,\bm{s},k)>\bm{0}$.

Finally, define the following central path in (HLP)
\begin{equation}
\begin{split}
    \mathcal{C}=\{&(\bm{y},\bm{x},\tau,\theta, \bm{s},k)\in \mathcal{F}_h^0:\\
    &\begin{pmatrix}
        X\bm{s}\\
        \tau k
        \end{pmatrix}
        =\frac{\bm{x}^T \bm{s}+\tau k}{n+1} 1_{(n+1)\times 1} \},
\end{split}
\end{equation}
and its neighbourhood
\begin{equation}
\begin{split}
    \mathcal{N}(\beta)=&\{(\bm{y},\bm{x},\tau,\theta, \bm{s},k)\in \mathcal{F}_h^0:\left|\left|\begin{pmatrix}
        X \bm{s}\\
        \tau k
        \end{pmatrix}-\mu 1_{(n+1)\times 1}\right|\right|\\ &
        \leq\beta \mu
        \text{ where }\mu=\frac{\bm{x}^T \bm{s}+\tau k}{n+1} \}.
        \label{central path neighbourhood}
\end{split}
\end{equation}
Then, theorem 5 of \cite{Predictor-Corrector} ensures that the central path lies in the feasibility region of (HLP).

In consequence, the algorithm proceeds as follows: start from an interior feasible point $\bm{v}^0=(\bm{y}^0,\bm{x}^0,\tau^0,\theta^0, \bm{s}^0,k^0)\in \mathcal{F}_h^0$. Then, recursively, form the following system of equations for variables $\bm{d}_{\bm{v}}=(\bm{d_y},\bm{d_x}, d_\tau, d_\theta, \bm{d_s}, d_k)$ and $t=0,1,...\in \mathbb{N}$:
\begin{subequations}
\begin{equation}
\begin{pmatrix}
           &+A &-\bm{b} &+\bar{\bm{b}}&  \\
       -A^T&  &+\bm{c} &-\bar{\bm{c}}& - 1\\
       +\bm{b}^T&-\bm{c}^T& & +\bar{z}& &-1\\
        -\bar{\bm{b}}^T&+\bar{\bm{c}}^T&-\bar{z}& & &
\end{pmatrix}
\begin{pmatrix}
\bm{d_y}\\
\bm{d_x}\\
d_\tau\\
d_\theta\\
\bm{d_s}\\
d_k
\end{pmatrix}
=
\begin{pmatrix}
\bm{0}\\
\bm{0}\\
0\\
0
\end{pmatrix}\label{eq11}
\end{equation}
\begin{equation}
    \begin{pmatrix}
    X^t \bm{d_s}+ S^t\bm{d_x}\\
    \tau^t d_k +k^t d_\tau
    \end{pmatrix}
    =\gamma^t \mu^t 1_{(n+1)\times 1} - \begin{pmatrix}
    X^t \bm{s}^t\\
    \tau^t k^t\end{pmatrix},
    \label{eq12}
\end{equation}
\end{subequations}
where $\gamma^t$ takes values 0 and 1 for even and odd steps respectively, starting in $t=0$. The linear system of equations can be written in matrix form as  $M^t \bm{d}_{\bm{v}^t}=\bm{f}^t$, i.e.
\begin{widetext}
\begin{equation}
\begin{blockarray}{ccccccc}
& m &  n &  1 &  1 & n & 1 \\
\begin{block}{c(cccccc)}
     m& 0 & A & -\bm{b} & \bar{\bm{b}} & 0  &  0  \\
     n & -A^T & 0 & \bm{c} & -\bar{\bm{c}} & -1 & 0 \\
     1 &\bm{b}^T & -\bm{c}^T & 0 &\bar{z} & 0 & -1 \\
     1& -\bar{\bm{b}}^T & \bar{\bm{c}}^T & -\bar{z} &0 & 0 & 0 \\
     n& 0 & S^t & 0 &  0 &  X^t & 0\\
     1& 0 &  0 & k^t & 0  &  0 & \tau^t\\
    \end{block}
    \end{blockarray}
    \qquad
    \begin{blockarray}{c}
    \\
    \begin{block}{(c)}
    \bm{d_y}\\
    \bm{d_x}\\
    d_\tau\\
    d_\theta\\
    \bm{d_s}\\
    d_k\\
    \end{block}
    \end{blockarray}\quad=\quad
    \begin{blockarray}{c}
\\
    \begin{block}{(c)}
    \bm{0}\\
    \bm{0}\\
    0\\
    0\\
    \gamma^t\mu^t 1_{n\times 1}-X^t \bm{s}^t\\
    \gamma^t\mu^t -\tau^tk^t\\
    \end{block}
    \end{blockarray}.
    \label{Matrix system}
\end{equation}
\end{widetext}
So, the main loop of the algorithm consists in performing the following steps iteratively:

\textit{Predictor step:} Solve \eqref{Matrix system} with $\gamma^t=0$ for $\bm{d}_{\bm{v}^t}$ where $\bm{v}^t=(\bm{y}^t,\bm{x}^t,\tau^t,\theta^t, \bm{s}^t,k^t)\in\mathcal{N}(1/4)$. Then find the biggest step length $\delta$ such that 
\begin{equation}
\bm{v}^{t+1}=\bm{v}^{t}+\delta \bm{d}_{\bm{v}^t}
\label{predictor sum}
\end{equation}
is in $\mathcal{N}(1/2)$, and update the values accordingly. Then $t \leftarrow t+1$.

\textit{Corrector step:} Solve \eqref{Matrix system} with $\gamma^t=1$ and set 
\begin{equation}
\bm{v}^{t+1}=\bm{v}^{t}+ \bm{d}_{\bm{v}^t} \label{corrector sum}
\end{equation}
that will be back in $\mathcal{N}(1/4)$. Update $t \leftarrow t+1$.

\subsection{\label{Termination}Termination}
Define a strictly self-complementary solution of (HLP) $\bm{v}^*=(\bm{y}^*,\bm{x}^*,\tau^*,\theta^*=0,\bm{s}^*,k^*)$ as an optimal solution to (HLP) that fulfills
\begin{equation}
    \begin{pmatrix}
    \bm{x}^*+\bm{s}^*\\
    \tau^*+k^*
    \end{pmatrix}
    >\bm{0}.
\end{equation}

Theorem 3 in \cite{Predictor-Corrector} tells us that if we have a strictly self-complementary solution to (HLP), then a solution to (LP) and (LD) exits whenever $\tau^*>0$, in which case $\bm{x}^*/\tau^*$ and $(\bm{y}^*/\tau^*,\bm{s}^*/\tau^*)$ are the solutions respectively.
On the other hand, if $\tau^*=0$ at least one of two things will happen: $\bm{c}^T\bm{x}^*<0$, meaning that (LD) is not feasible, or $-\bm{b}^T\bm{y}^*<0$ in which case (LP) is not feasible.

The loop from the previous section will run over $t$ until one of the following two criteria are fulfilled: For $\epsilon_1,\epsilon_2,\epsilon_3$ small numbers, either 
\begin{subequations}
    \begin{equation}
    \begin{split}
        (\bm{x}^t/\tau^t)^T(\bm{s}^t/\tau^t)\leq \epsilon_1 \text{ and }\\
        (\theta^t/\tau^t)||(\bar{\bm{b}}^T,\bar{\bm{c}}^T)||\leq \epsilon_2, \label{termination condition 1}        
    \end{split}
    \end{equation}
or 
    \begin{equation}
        \tau^t \leq \epsilon_3. \label{termination condition 2}
    \end{equation}
\end{subequations}
We can see that the two equations in \eqref{termination condition 1} are related to the dual gap being 0, and $\theta^*=0$ (needed conditions for the solution to be optimal); supposing $\tau^*>0$. The equation \eqref{termination condition 2} is the procedure to detect $\tau^*=0$. $\epsilon_1$ and $\epsilon_2$ should therefore be chosen taking into account the precision we are seeking in the optimality of the solution, and the error our calculations will have. In particular, $\epsilon_1$ and $\epsilon_2$ can be taken to be the target error of the algorithm, $\epsilon$.

To get to this point we will have to iterate up to $O(L\bar{t}\sqrt{n})$ times, with $\bar{t}=\max[\log((\bm{x}^0)^T( \bm{s}^0)/(\epsilon_1\epsilon_3^2)), \log(||(\bar{\bm{b}}^T,\bar{\bm{c}}^T)||/\epsilon_2\epsilon_3)]$.

If the termination is due to condition \eqref{termination condition 2}, then we know that there is no solution fulfilling $||(\bm{x},^T\bm{s}^T)||\leq 1/(2\epsilon_3)-1$. Therefore one should choose $\epsilon_3$ small enough so that the region we are exploring is reasonable. We will then consider, following \cite{Predictor-Corrector}, that either (LP) or (LD) are infeasible or unbounded.

However, if termination is due to \eqref{termination condition 1}, denote by $ \zeta^t$ the index set $\{j\in 0,...,n: \bm{x}_j^t\geq  \bm{s}^t_j\}$. Let also $B$ the columns of $M^t$ such that their index is in $ \zeta^t$, and the rest by $C$. 

\textit{Case 1:} If $\tau^t\geq k^t$ solve for $\bm{y}, \bm{x}_B, \tau$
\begin{subequations}    \label{termination 1}
\begin{equation}
\label{termination 1a}
     \min_{\bm{y},\bm{x}_B,\tau} ||\bm{y}^t-\bm{y}||^2+||\bm{x}^t_B-\bm{x}_B||^2+(\tau^t-\tau)^2   
\end{equation}
such that
\begin{equation}
\label{termination 1b}
    B\bm{x}_B-\bm{b}\tau=0; \quad -B^T\bm{y}+\bm{c}_B\tau=0; \quad \bm{b}^T\bm{y}-\bm{c}^T_B\bm{x}_B=0;
\end{equation}
\end{subequations}

\textit{Case 2:} If $\tau^t<k^t$ and we solve for $\bm{y}, \bm{x}_B,$ and $k$ from
\begin{subequations}
    \begin{equation}
    \min_{\bm{y},\bm{x}_B,k} ||\bm{y}^t-\bm{y}||^2+||\bm{x}^t_B-\bm{x}_B||^2+(k^t-k)^2 \label{termination 2a}
    \end{equation}
    such that
    \begin{equation}
    B\bm{x}_B=0; \quad -B^T\bm{y}=0; \quad \bm{b}^T\bm{y}-\bm{c}^T_B\bm{x}_B-k=0. \label{termination 2b}
    \end{equation}
    \label{termination 2}
\end{subequations}
The result of either of these two calculations will be the output of our algorithm, and the estimate of the solution of the (HLP) problem. In particular, $\bm{x}$ will be the calculated $\bm{x}_B$ in the least square projection together with $\bm{x}_C$, and $\bm{y}$ will be the calculated $\bm{y}$ again in the least square projection. Calculating the solution to (LP) and (LD) is then straightforward: $\bm{x}^*/\tau^*$ and $(\bm{y}^*/\tau^*,\bm{s}^*/\tau^*)$ respectively.

\section{\label{sec:The quantum algorithm}The quantum algorithm}
The aim of this section is to explain how the Quantum Linear System Algorithm (QLSA) can help us efficiently run this algorithm, in the same spirit of, for example, \cite{FEM} solving the problem of the Finite Element Method. This is due to the fact that solving \eqref{Matrix system} is the most computationally expensive part of each step for large matrices. 

To solve this system we propose leveraging the use of a technique called block encoding or qubitization, introduced in \cite{qubitization}. Using such technique, which we shall expand further, we have the following result (theorem 34 from \cite{block-encoding_1}).

\begin{Theorem} \label{QLSA theorem}
\cite{block-encoding_1}:
Let $M$ be an $n' \times n'$ Hermitian matrix (if the matrix is not Hermitian it can be included as a submatrix of a Hermitian one) with condition number $\kappa$, Frobenius norm $||M||_F=\sqrt{\sum_{ij}|M_{ij}|^2}$ and spectral norm $||M||\leq 1$. Let $\bm{f}$ be an $n'$-dimensional unit vector, and assume that there is an oracle $\mathcal{P}_f$ which produces the state $\ket{f} = \bm{f}/||\bm{f}||$ in time $T_f$. Let also $M$ be encoded in the quantum accessible data structure indicated in \ref{sec:state_preparation}. Let
\begin{equation}
    \bm{d_v}=M^{-1}\bm{f}, \qquad \ket{d}=\frac{\bm{d_v}}{||\bm{d_v}||}.\label{Md=f}
\end{equation}
Then, 
\begin{enumerate}
    \item We can prepare state $\ket{d}$ in time complexity
    \begin{equation}
        \Tilde{O}((||M||_F + T_f)\kappa \text{ poly}\log(n'\epsilon^{-1})).
    \end{equation}
    \item We can get an $\epsilon$-multiplicative estimate of $||M^{-1}\ket{f}||$ in time
    \begin{equation}
        \Tilde{O}((||M||_F + T_f)\kappa \epsilon^{-1} \text{ poly}\log(n'\epsilon^{-1})).        
    \end{equation}
\end{enumerate}
\end{Theorem}
Proof omited. \qed

In our case the variable $n'$ is the size of the matrix of \eqref{Matrix system}, that is $n'=2(m+2n+3)$, the $2$ coming from symmetrisation as in the HHL algorithm. For spectral norm bounded matrices, $||M||\leq C$ constant, $||M||_F = O(\sqrt{n'})$ \cite{DenseHHL}. Thus, the time complexity of running the algorithm would be $O(\sqrt{n'})$. We are also assuming $m=O(n)$. Notice that since $n$ appears in the number of iterations but $m$ does not, it is convenient to set $m\geq n$ by exchanging the primal and dual problems if needed.

An alternative could be to use a combination of the Hamiltonian simulation for dense matrices of \cite{Dense_hamiltonian_simulation} with the Quantum Linear System Algorithm from \cite{QLSAchilds}, that would result in slightly different complexity factors.

Let us know study how to integrate this algorithm within the Interior-Point algorithm.

\subsection{\label{sec:condition number}The condition number $\kappa$.}
We have seen that the QLSA is linear in $\kappa$. Therefore it is important to check that $\kappa$ is as low as possible.

However, preconditioning a dense matrix is much more complicated than a sparse matrix. In fact, we are not aware of any method that allows us to do it without incurring in expensive computations in the worst case. For example, the method proposed in \cite{SPAI-QLSA} is only useful for sparse matrices.

Thus, as preconditioning does not seem possible, we might attempt setting an upper bound to $\kappa$ for all steps of the iteration, taking into account that only a small part of the matrix $M^t$ depends on $t$. The entries that depend on $t$ are the $n+1$ last rows, $2(n+1)$ entries, see \eqref{Matrix system}. However, if we try doing that we will see that even if it is possible to upper bound the maximum singular value $\sigma_{\max} (M^t)$ knowing the entries of the last rows, we cannot see a way to lower bound $\sigma_{\min} (M^t)$, so we cannot bound the condition number. 

In conclusion, we have not been able to bound the condition number from the start, so we have to rely on a rather unknown upper bound $\bar{\kappa}$. We remark that this shortcoming of the algorithm is common to both our algorithm, and the Predictor-Corrector if we substituted QLSA by other iterative methods.

\subsection{\label{sec:state_preparation}Quantum state preparation and quantum-accessible data structure}

In order to prepare quantum states there are many options that include \cite{State_prep_1,State_prep_grover,State_prep_2}. However we are interested here in some method that can allow us to prove some quantum advantage for the case. 

So, in order to do this, we introduce the method of \cite{block-encoding_1}, which additionally we will need in order to apply the QLSA. 

\begin{Theorem} \label{quantum accessible data structure}
 \cite{kerenidis2017recommendation, block-encoding_1}:
Let $M\in \mathbb{R}^{n'\times n'}$ be a matrix. If $w$ is the number of nonzero entries, there is a quantum accessible data structure of size $O(w\log^2(n'^2))$, which takes time $O(\log (n'^2))$ to store or update a single entry. 
Once the data structure is set up,  there are quantum algorithms that can perform the following maps to precision $\epsilon^{-1}$ in time $O(\text{poly}\log(n'/\epsilon))$:
\begin{equation}
    U_\mathcal{M}:\ket{i}\ket{0}\rightarrow \frac{1}{||M_{i\cdot}||}\sum_{j}M_{ij}\ket{ij};
    \label{mathcal M}
\end{equation}
\begin{equation}
    U_\mathcal{N}:\ket{0}\ket{j}\rightarrow \frac{1}{||M||_F}\sum_i ||M_{i \cdot}|| \ket{ij};
    \label{mathcal N}
\end{equation}
where $||M_{i \cdot}||$ is the $l_2-$norm of row $i$ of $M$.
This means in particular that given a vector $f$ in this data structure, we can prepare an $\epsilon$ approximation of it, $1/||v||_2 \sum_i v_i \ket{i}$, in time $O(\text{poly}\log(n'/\epsilon))$.
\end{Theorem}

Proof: 
To construct the classical data structure, create $n'$ trees, one for each row of $M$. Then, in leaf $j$ of tree $B_i$ one saves the tuple ($M_{ij}^2$, sgn($M_{ij}$)). Also, intermediate nodes are created (that join nearby branches) so that node $l$ of tree $B_i$ at depth $d$ contains the value
\begin{equation}
    B_{i,l}=\sum_{j_1,...,j_d = l}M_{ij}^2.
\end{equation}
Notice that $j_1,...,j_d$ is a string of values $0$ and $1$, as is $l$. The root node contains the value $||M_{i\cdot}||^2$.

An additional tree is created taking the root nodes of all the other trees, as the leaves of the former. One can see that the depth of the structure is polylogarithmic on $n'$, and so a single entry of $M$ can be found or updated in time polylogarithmic on $n'$.

Now, to apply $U_\mathcal{M}$, we perform the following kind of controlled rotations
\begin{equation}
\begin{split}
    &\ket{i}\ket{l}\ket{0...0}\rightarrow\\ &\ket{i}\ket{l}\frac{1}{\sqrt{B_{i,l}}}\left(\sqrt{B_{i,2l}}\ket{0}+\sqrt{B_{i,2l+1}}\ket{1}\right)\ket{0...0},
\end{split}
\end{equation}
except for the last rotation, where the sign of the leaf is included in the coefficients. It is simple to see that $U_\mathcal{N}$ is the same algorithm applied with the last tree, the one that contains $||M_{i\cdot}||$ for each $i$. Finally, for a vector, we have just one tree, and the procedure is the same. \qed

One may worry about two things: the first is that setting up the database might take too long, since our matrices are dense. However, notice that in $M^t$ only $O(n+m)$ entries depend on $t$, so the rest can be prepared at the beginning of the algorithm with an overall time complexity of $O((n+m)^2)$, up to polylogarithmic factors. This is the same complexity as the overall algorithm when matrices are spectrally bounded.

On the other hand twice per iteration one must update the entries in the last $n+1$ rows of $M^t$, and prepare the data structures for the preparation of the quantum states, which will take time $O(n+m)$, but that is fine since the work complexity on $n+m$ is the same as needed to read out the result, and so will not add any complexity to the result, and it has to be done just once for each linear system of equations.

Finally, preparing the states $\ket{f}$ themselves comes at a polylogarithmic cost in both $n+m$ and $\epsilon$, so we do not need to care about it. This quantum state preparation protocol seems particularly useful when we want to solve the same linear system of equations multiple times to read out the entire solution.

\subsection{\label{sec:Readout} Readout of the solution of QLSA.}

In the same way that we need some procedure to prepare the quantum state that feeds in the QLSA, we need some way to read out the information in $\ket{d}$, defined as in equation \eqref{Md=f}. 

This is a tomography problem where the solution is a pure state with real amplitudes. Thus, one may be inclined to use Amplitude Estimation to read each of the entries. However, this has the problem that entries of large normalized vectors will be small, making it costly to determine each amplitude with the additive precision provided by Amplitude Estimation.

So, instead of that procedure, we propose using the tomography algorithm explained in section 4 of \cite{kerenidis2018quantum}. The complexity of the tomography is $O(n'\epsilon^{-2})$. Theorem 4.3 in \cite{kerenidis2018quantum} proves that if $\ket{\Tilde{d}}$ is the output of the tomography algorithm \ref{Tomograpy}, then $||\ket{\Tilde{d}}- \ket{d}||_2\leq \sqrt{7}\epsilon$ with probability greater or equal to $1-1/n'^{0.83}$. Notice that theorem \ref{QLSA theorem} allows us to recover the norm of the solution. The global sign of the solution might be recovered multiplying a single row of $A$ with the solution, and comparing the result against the corresponding entry of $\bm{f}^t$.

\begin{figure}
\begin{algorithm}[H]
\caption{Vector tomography \cite{kerenidis2018quantum}}\label{Tomograpy}
\begin{algorithmic}[1]
\Procedure{Vector tomography}{}
\BState \textbf{Amplitude estimation}
\State Prepare $N=\frac{36 n' \log n'}{\epsilon^2}$ copies of $\ket{d}$, and measure them, obtaining $n_i$ times the basis state $\ket{i}$. Define amplitudes $\sqrt{p_i} = \sqrt{n_i/N}$.
\State Save the vector $\ket{p}=\sum_i \sqrt{p_i}\ket{i}$ in the quantum accessible data structure from theorem \ref{quantum accessible data structure}.

\BState \textbf{Sign estimation (Swap Test)}
\State  Create $N=\frac{36 n' \log n'}{\epsilon^2}$ copies of the state $\frac{1}{\sqrt{2}}\ket{0}\ket{d}+ \frac{1}{\sqrt{2}}\ket{0}\ket{d}$.
\State Apply a Hadamard gate to the control qubit in each copy of the state in the previous step. This prepares
\begin{equation}
    \frac{1}{2}\sum_i [(d_i + \sqrt{p_i})\ket{0,i} + (d_i - \sqrt{p_i})\ket{1,i}].
\end{equation}
\State Measure the states in the computational basis, obtaining frequencies $n(b,i)$, where $b = 0,1$.
\State Set the sign $\sigma_i = +1$ if $n(0,i) > 0.4 p_i N$. Else, $\sigma_i = -1$.
\State Output the classical vector with entries $\sigma_i \sqrt{p_i}$.
\EndProcedure
\end{algorithmic}
\end{algorithm}
\end{figure}

\subsection{\label{sec:QLSA}Block encoding and Quantum Linear System Algorithm (QLSA)}

In this subsection let us briefly review some of the results that are needed for the Quantum Linear system algorithm. In the first place, let us review what we mean by block encoding (definition 3 in \cite{block-encoding_1}).

\begin{Definition}
Block encoding \cite{block-encoding_1} Given a $s$-qubit operator $A$, the unitary operator $U$ is an $(a,\alpha,\epsilon)$-block encoding of $A$ when 
\begin{equation}
    ||A - \alpha(\bra{0}^{\otimes a}\otimes 1)U(\ket{0}^{\otimes a}\otimes 1)||\leq \epsilon.
\end{equation}
That means that one can write
\begin{equation}
    U =\begin{pmatrix}
    A/\alpha & \cdot \\
    \cdot & \cdot
    \end{pmatrix}.
\end{equation}
\end{Definition}

Block encodings are useful because they allow for a variety of linear algebra techniques including multiplication of block encodings, exponentiation to positive and negative powers, and Hamiltonian simulation; efficiently. For more information we refer the reader to \cite{block-encoding_1,block-encoding_2}.

Another important algorithm we are using is Variable Time Amplitude Amplification \cite{Ambainis} and Variable Time Amplitude Estimation \cite{block-encoding_1}. These algorithms were developed to reduce the quadratic complexity in the condition number $\kappa$ in the original HHL algorithm \cite{HHL}. In that algorithm the quadratic complexity appears because of the $\kappa$ complexity in phase estimation and the $\kappa$ cost in amplitude amplification (or estimation).

The insight of \cite{Ambainis} was to realise that for the eigenvalues $\lambda$ for which the phase estimation was costly ($\lambda$ small) the cost of amplitude amplification is low, and viceversa. Thus, he proposed variable time algorithms, that allow to stop some branches of the algorithm before others. Since the more expensive branches require less Amplitude Amplification, they stop earlier. The result is a reduction of the cost to linear in $\kappa$. For more information on this complex algorithm we refer the reader to those references. These two elements are the key ideas needed to obtain the result of theorem \ref{QLSA theorem}.

\subsection{\label{sec:Quantum termination}On quantizing the termination.}

If there exist a feasible and optimal solution, we have seen that the loop should terminate with either procedures \eqref{termination 1} or \eqref{termination 2}. However it is unclear how to carry out this minimization. What we know is that it can be efficiently calculated, or substituted by any more modern and efficient method if found. 

The cost of carrying out this termination by classical procedures should be not too big. In fact, according to \cite{ye_termination} the overall cost is around that of one iteration of the main loop. 

However, we can also propose a quantum method to finish this. It would consist on using a small Grover subroutine \cite{Grover} to find all solutions of \eqref{termination 1b} or \eqref{termination 2b} in a small neighbourhood of the latest calculated point. After that, without reading out the state, one could apply the algorithm described in \cite{GroverMin} to calculate the one with the smallest distance to the calculated point, as in \eqref{termination 1a} or \eqref{termination 2a}.

\subsection{\label{sec:complexity}Complexity}
In this section we will indicate the complexity of our algorithm against other algorithms that can be used to solve Linear Programming problems. In particular, in table \ref{tabla comparativa complejidad} we compare against the same Predictor-Corrector algorithm but using one iterative Classical Linear System Algorithm (conjugate gradient descent \cite{conjugate_gradient}), two exact classical methods (like Gauss or Cholesky decomposition, or the optimal exact algorithm \cite{dense_exact}), and against the recent quantum algorithms proposed by Brand\~ao and Svore \cite{Brandao-Svore}, and Keredinis and Prakash \cite{kerenidis2018quantum} for solving Semi Definite Programming problems, a more general class of problems than those studied here (Linear Programming).

Firstly, we must take into account  that, as we are using the Predictor-Corrector algorithm \cite{Predictor-Corrector}, that means by construction $O(\sqrt{n}L)$ iterations of the main loop. For dense problems (as those we are considering), we should also take into account the complexity of solving two Linear Systems of Equations. The QLSA we are using is described in \cite{block-encoding_1}, with complexity $O(\overline{||M||_F}\bar{\kappa} \text{poly}\log((n+m)\epsilon^{-1}))$. 
In contrast, the fastest comparable Classical Linear System Algorithm is the conjugate gradient method \cite{conjugate_gradient}, which has time complexity $O((n+m)^2\bar{\kappa}\log(\epsilon^{-1}))$ for general (not symmetric, positive semidefinite) dense matrices.

But we also have to take into account other procedures. Those are: the preparation of quantum states has work complexity $O(n+m)$ if we take into account the preparation of the classical data structure, and the tomography requires $O((n+m)\epsilon^{-2})$ complexity, multiplied by the complexity of QLSA.

In general we have for our algorithm a runtime of $O(L\sqrt{n}(n+m)\overline{||M||_F}\bar{\kappa}\epsilon^{-2})$, where each iteration comes at a runtime cost of $O((n+m)\overline{||M||_F}\bar{\kappa}\epsilon^{-2})$, up to polylogarithmic terms. All of this is quantum work, since the only classical operations we perform are multiplicating the vectors needed to find $\delta$ in the Predictor step, and recalculating and updating the data-base for $M^t$ and $f^t$ in each round. In the next section, \ref{sec:Failure}, we will see that some times a small number of steps of a gradient descent will be necessary after the Corrector steps, with cost $O(n+m)\text{poly}\log (n+m)$, which we already had. Finally, an additional inexpensive classical postprocessing step after the Predictor step will be needed to ensure the same convergence guarantees as the original Predictor Corrector algorithm.

It is also remarkable to mention that, thanks to the de-quantization algorithm of Ewin Tang \cite{tang2018quantum}, it is possible to solve linear systems of equations (and therefore use our Interior-Point algorithm) in work complexity $O(||M||_F^6 k^6 \kappa^{16} \epsilon^{-6})$. Therefore, this classical algorithm is only useful if the rank of the matrix $k$ is low compared to $O(n+m)$ \cite{chia2018quantum,arrazola2019quantum}. However notice that we have not made any assumption about the rank of the matrix $A$, and for Linear Programming problems we do not expect this to be the case in general.

\subsection{\label{sec:Failure}Probability of failure}

Finally, we want to analyze the probability of failure of the algorithm. The reason for this is because the classical Predictor-Corrector algorithm assumes exact arithmetic, and we have to take care of the error $\epsilon$. What is more, the time complexity of the algorithm is quadratic on the associated precision $\epsilon^{-1}$, so it is computationally expensive to reach high precision. The failure of the algorithm may happen because we get out of $\mathcal{N}(\beta)$ in one of the steps. Let us now analyze if this is in fact possible.

In the Predictor steps we can state that the failure is not possible. This is because we are moving from $\mathcal{N}(1/4)$ to $\mathcal{N}(1/2)$ where we classically calculate $\delta$ such that this step is performed correctly. Therefore there is no chance of failure here, since in the worst case we can always find $\delta$ small enough such that in \eqref{predictor sum} $\bm{v}^{t+1}\in\mathcal{N}(1/2)$ if $\bm{v}^{t}\in\mathcal{N}(1/4)$.

In the Corrector steps the problem is different since now we are moving from $\mathcal{N}(1/2)$ to $\mathcal{N}(1/4)$ and there is no parameter we can tune. Therefore, one could think that the conditions of a complexity $O(\epsilon'^{-2})$ on the error parameter of the quantum subroutine (that forces us to have a loose error) and the possibility of a corrector step getting out of $\mathcal{N}(1/4)$ (that forces a high precision) could be incompatible.

In this section we will see that from a naive point of view they could seem incompatible, but in reality we can avoid this problem. To do this we will introduce a very simple procedure that allows us to ensure that even if we have a relatively high error, we still end up inside $\mathcal{N}(1/4)$.

The first idea one may have to check whether we end up in $\mathcal{N}(1/4)$, is  to look at lemma 3 in \cite{Predictor-CorrectorI}, in which \cite{Predictor-Corrector} is based. If we analyze the details of the proof we can see that in fact the exact solution of the Corrector Step is not only in $\mathcal{N}(1/4)$ but also in $\mathcal{N}(1/4\sqrt{2})$. This means that we need to lower the error sufficiently so that if the exact arithmetic result is in $\mathcal{N}(1/4\sqrt{2})$, then the approximate solution is in $\mathcal{N}(1/4)$. In the appendix \ref{sec:epsilon'} we see that in fact this would require very low error condition, something that the quantum subroutine cannot provide in reasonable time complexity. There we derive the full calculation for completeness, even if they are a bit tedious. In particular the precision needed would be $\epsilon'^{-1}= O(n^{\text{poly}\log n})$. The reason for the $n$ dependence on the precision is related to the fact that the error $\epsilon'$ affects each coordinate of $\bm{x}^t$ or $\bm{s}^t$, and there are $O(n)$ of them.

To solve this problem let us introduce a different strategy. Choose a low precision on the error for the corrector step $\epsilon'^{-1} = O(n^{1/a})$, with $a$ sufficiently large. This is `cheap' in terms of the complexity of the quantum subroutine for $a$ relatively large. We will now shift the result slightly ($\epsilon''$ close) but in such a way that the new point fulfils $||X\bm{s}-\mathbf{1}\mu||\leq \beta\mu$, thus introducing an error of size $\epsilon''$ in each coordinate of the result. To do that we can perform one step of gradient descent (or a small number of them). Therefore we shift each coordinate of the output of the Corrector step at most $\epsilon''$. How large is that $\epsilon''$? Intuitively one can think that if the error of the Corrector step is at most $\epsilon'$ for each coordinate, then $\epsilon'' = O(\epsilon')$, which we already had anyway. As we calculate in appendix \ref{sec:epsilon''} this is in fact true and performing gradient descent with such $\epsilon''$ is sufficient to end up inside $\mathcal{N}(1/4)$. We want to emphasise that the legitimacy of this procedure rests on the fact that the precision on the Corrector step is only important because it could cause the algorithm to fail by ending outside $\mathcal{N}(1/4)$, apart from affecting the final error of the algorithm, $\epsilon$. Finally, note that this procedure has time complexity $O(n+m)$ as calculated in appendix \ref{sec:epsilon''}, so it adds no additional complexity to the algorithm. In appendix \ref{sec:convergence} we also see that imposing \eqref{Predictor shift}, which has no complexity consequences, we can prove that our modification does not affect the convergence of our algorithm, and so the number of iterations of our algorithm is the same as the original Predictor-Corrector.

The conclusion of the previous is that, in order to comply with both the requirements of having a low complexity in $\epsilon'$ in the quantum subroutine and still ending up in $\mathcal{N}(1/4)$ in the end of the corrector step, we can $O(\epsilon')$-shift the output of the quantum subroutine and make it comply with the second requirement while $\epsilon'$ is sufficiently large. This only imposes is that the final precision we want from our algorithm ($\epsilon^{-1}$) cannot be greater than the one induced by the error $\epsilon'$ that the Corrector step and `shifting' procedure has in the worst case. Or, in other words, $\epsilon \geq O(\epsilon') = O(n^{-1/a})$. This should nevertheless pose no problem since we assume that in general we do not ask for a greater precision in each variable for a larger problem, what means that $\epsilon = O(1)$ in the variable $n$. In other words, even if the target error $\epsilon$ for each variable is small, it will have a priori no relation with the number of variables $n$ or constraints $m$. It is reasonable to assume this since in general one is interested in a given precision for each variable, independent of their number.

Another way our algorithm could fail is because in each step we saw that the tomography has a failure probability of $1/(3+2n+m)^{0.83}$. Let us analyze if that is a problem. First notice that after each step we can easily check that the solution is correct up to $\epsilon$ precision. For example, using the quantum accessible data structure we can prepare the trial solution, and perform a swap test to check that the solution is $\epsilon$-close to the actual solution. The swap test may be combined with Amplitude Estimation and the median lemma from \cite{nagaj2009fast} to yield exponentially small failure probability. This means that if we perform $c$ attempts at each step, the probability of failure in each iteration decreases to $1/(3+2n+m)^{0.83 c}$.

So, we have to choose a small constant $c$ such that after $O(\sqrt{n}L)$ iterations the total probability of failure $(\sqrt{n}L)/(3+2n+m)^{0.83 c} \ll 1$. Since in many cases the number of iterations will be $O(\sqrt{n}\log n)$ taking $c = 1$ should do, but in general $L = O(mn)$, so we can take $c = 4> 2.5/0.83$. In conclusion, the probability of failure of the tomography algorithm should pose no problem.

Finally, in order to decrease the final error of the algorithm even further, in practice it could be a good idea to perform a small (constant number) of iterations classically (doing this will not affect the theoretical complexity). Theoretically though, this does not give any guarantee of success, since the number of additional iterations would be $O(\sqrt{n}L(\bar{t}_{\epsilon_a}-\bar{t}_{\epsilon_b}))$, when we want to lower the error from $\epsilon_b$ to $\epsilon_a$. Even if the difference is small, there is a dependence on $\sqrt{n}$ that would increase the complexity of the algorithm making it comparable to the complexity of classical algorithms.

We already explained in section \ref{Termination} that the number of iterations should be $\bar{t}=\max[\log((\bm{x}^0)^T( \bm{s}^0)/(\epsilon_1\epsilon_3^2)), \log(||(\bar{\bm{b}}^T,\bar{\bm{c}}^T)||/\epsilon_2\epsilon_3)]$. In particular, if any one supposes every entry of $\bm{x}^0$, $\bm{s}^0$, $\bm{b}$ and $\bm{c}$ to be of order $O(1)$, then $\bar{t}_a = \log (O(n+1)/\epsilon^2_a)$ and similarly for $\bar{t}_b$ and $\epsilon_b$. This would mean that to lower the error from $\epsilon_b$ to $\epsilon_a$, one would need
\begin{equation}
    \bar{t}_a-\bar{t}_b = \log \frac{n+1}{\epsilon^2_a}-\log \frac{n+1}{\epsilon^2_b} = \log \frac{\epsilon^2_b}{\epsilon^2_a}=O(n^{-1/2}),
\end{equation}
what implies that
\begin{equation}
    \epsilon_a = \epsilon_b e^{-1/2\sqrt{n}}.
\end{equation}

Since $\lim_{n \rightarrow \infty} e^{-1/2\sqrt{n}} = 1$, we would need the final error $\epsilon_a$ to be very similar to the one we achieve with the quantum subroutine $\epsilon_b$. Or, with the notation we are using: $\epsilon_a = O(\epsilon_b) = O(n^{-1/a})$.

\begin{widetext}
\begin{figure*}
    \centering
    \includegraphics[width=620pt, angle=90]{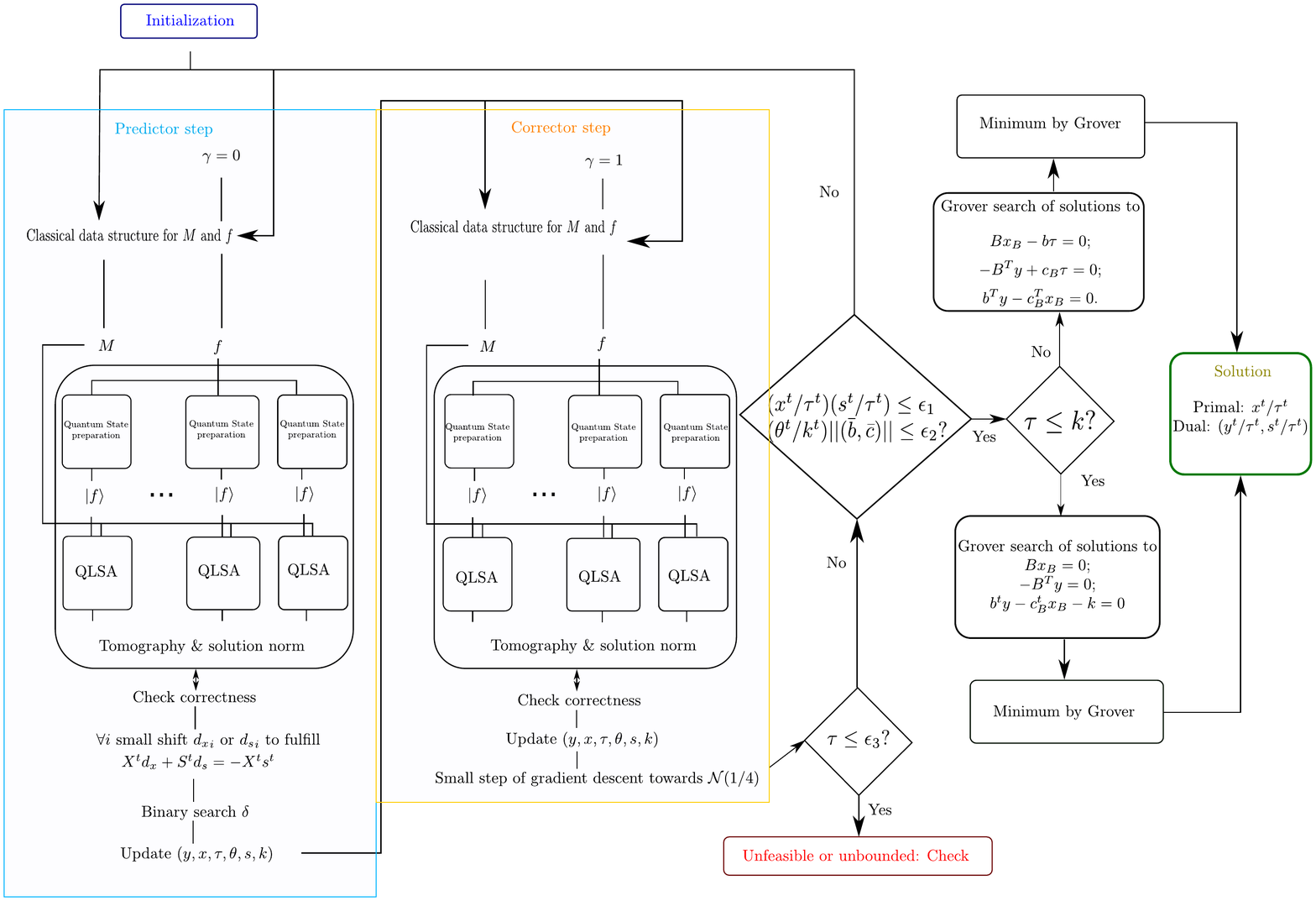}
    \caption{Flow chart of the algorithm}
    \label{fig:flow_chart}
\end{figure*}
\end{widetext}

To summarize all the components in our quantum Predictor-Corrector algorithm and the interrelations among them, we show a diagram in Fig. \ref{fig:flow_chart} in the form of a flow chart of actions from the initialization to the termination of the quantum algorithm  providing the solution to the given (LP) and (LD) problems in \eqref{LP} and \eqref{LD}.

\pagebreak

 \section{Conclusions}
 
 Quantization of Linear Programming problems thus far have been achieved by using multiplicative weight methods as in the pioneering work of Brand\~ao and Svore for Semidefinite Programming (SDP) problems \cite{Brandao-Svore}, which are more general than Linear Programming problems. In this work,  we have enlarged the range of applicability of quantum algorithms for Linear Programming problems by using Interior Point methods instead. 
Specifically, our quantum algorithm relies on a type of Interior Point algorithm known as the Predictor-Corrector method that is very well behaved with respect to the feasibility,  optimality conditions of the output solution and the iteration complexity.

The core of our quantum Interior Point algorithm is the application of block encoding techniques as a Quantum Linear System Algorithm \cite{block-encoding_1} to an auxiliary system of equations that comprises an homogeneous self-dual primal-dual problem associated to the original Linear Programming problem. This is the basis of the Predictor-Corrector method, from which many of its good properties derive. In particular, the iteration complexity of the classical part scales as the square root of the size $n$ of the cost function.
Then, the advantage of the quantum part of the Predictor-Corrector algorithm amounts to a faster solution of the linear system of equations, with complexity $O((n+m)\sqrt{n+m})$ including the readout process, as can be seen in table \ref{tabla comparativa complejidad}.

Hence, this quantum Predictor Corrector algorithm is an hybrid algorithm, partially classical, partially quantum.
Applying the QLSA is not an easy task if we want to achieve a clear advantage. These algorithms come with several shortcomings, some of which have been recently overcome \cite{SPAI-QLSA} for sparse linear systems. Also, even though the solution to the system of linear equations can be obtained in a quantum state, then it is not easy to extract all the information provided by the solution. One has to be satisfied by obtaining partial information from the encoded solution such as an expectation value of interest or a single entry of the vector solution.
Nevertheless this does not stop us from obtaining a polynomial quantum advantage in the number of variables of the problem $n$, if the matrix is dense, well-conditioned, with $m=O(n)$, and with a constant-bounded spectral norm.

\section*{Acknowledgements}
We acknowledge financial support from the Spanish MINECO grants  FIS2015-67411P, and the CAM research consortium QUITEMAD-CM, Grant No.S2018/TCS-4342.  The research of M.A.M.-D. has been partially supported by the U.S. Army Research Office through Grant No. W911NF-14-1-0103. P. A. M. C. thanks the support of a FPU MECD Grant.

\bibliographystyle{ieeetr}

\begin{thebibliography}{10}

\bibitem{Predictor-Corrector}
Y.~Ye, M.~J. Todd, and S.~Mizuno, ``An $o(\sqrt{n}l)$-iteration homogeneous and
  self-dual linear programming algorithm,'' {\em Mathematics of Operations
  Research}, vol.~19, no.~1, pp.~53--67, 1994.

\bibitem{block-encoding_1}
S.~Chakraborty, A.~Gily{\'e}n, and S.~Jeffery, ``The power of block-encoded
  matrix powers: improved regression techniques via faster hamiltonian
  simulation,'' {\em arXiv preprint arXiv:1804.01973}, 2018.

\bibitem{Nering-Tucker}
E.~D. Nering and A.~W. Tucker, {\em Linear Programs \& Related Problems: A
  Volume in the Computer Science and Scientific Computing Series}.
\newblock Elsevier, 1992.

\bibitem{Padberg}
M.~Padberg, {\em Linear optimization and extensions}, vol.~12.
\newblock Springer Science \& Business Media, 2013.

\bibitem{Murty}
K.~G. Murty, {\em Linear programming}, vol.~60.
\newblock Wiley New York, 1983.

\bibitem{Russell-Norvig}
S.~J. Russell and P.~Norvig, {\em Artificial intelligence: a modern approach}.
\newblock Pearson Education Limited,, 2016.

\bibitem{MRT}
M.~Mohri, A.~Rostamizadeh, and A.~Talwalkar, {\em Foundations of machine
  learning}.
\newblock MIT press, 2012.

\bibitem{Vandernberghe-Boyd}
L.~Vandenberghe and S.~Boyd, ``Semidefinite programming,'' {\em SIAM review},
  vol.~38, no.~1, pp.~49--95, 1996.

\bibitem{Todd}
M.~J. Todd, ``Semidefinite optimization,'' {\em Acta Numerica}, vol.~10,
  pp.~515--560, 2001.

\bibitem{Laurent-Rendl}
M.~Laurent and F.~Rendl, {\em Semidefinite programming and integer
  programming}.
\newblock Centrum voor Wiskunde en Informatica, 2002.

\bibitem{deKlerk}
E.~De~Klerk, {\em Aspects of semidefinite programming: interior point
  algorithms and selected applications}, vol.~65.
\newblock Springer Science \& Business Media, 2006.

\bibitem{Preskill}
J.~Preskill, ``Quantum computing in the nisq era and beyond,'' {\em arXiv
  preprint arXiv:1801.00862}, 2018.

\bibitem{Ions}
D.~Nigg, M.~Mueller, E.~A. Martinez, P.~Schindler, M.~Hennrich, T.~Monz, M.~A.
  Martin-Delgado, and R.~Blatt, ``Quantum computations on a topologically
  encoded qubit,'' {\em Science}, p.~1253742, 2014.

\bibitem{SCQ1}
R.~Barends, J.~Kelly, A.~Megrant, A.~Veitia, D.~Sank, E.~Jeffrey, T.~C. White,
  J.~Mutus, A.~G. Fowler, B.~Campbell, {\em et~al.}, ``Superconducting quantum
  circuits at the surface code threshold for fault tolerance,'' {\em Nature},
  vol.~508, no.~7497, p.~500, 2014.

\bibitem{SCQ2}
A.~D. C{\'o}rcoles, E.~Magesan, S.~J. Srinivasan, A.~W. Cross, M.~Steffen,
  J.~M. Gambetta, and J.~M. Chow, ``Demonstration of a quantum error detection
  code using a square lattice of four superconducting qubits,'' {\em Nature
  communications}, vol.~6, p.~6979, 2015.

\bibitem{Preskill2}
J.~Preskill, ``Quantum computing and the entanglement frontier,'' {\em arXiv
  preprint arXiv:1203.5813}, 2012.

\bibitem{Aaronson-Arkhipov}
S.~Aaronson and A.~Arkhipov, ``The computational complexity of linear optics,''
  in {\em Proceedings of the forty-third annual ACM symposium on Theory of
  computing}, pp.~333--342, ACM, 2011.

\bibitem{Shor}
P.~W. Shor, ``Polynomial-time algorithms for prime factorization and discrete
  logarithms on a quantum computer,'' {\em SIAM review}, vol.~41, no.~2,
  pp.~303--332, 1999.

\bibitem{Grover}
L.~K. Grover, ``Quantum mechanics helps in searching for a needle in a
  haystack,'' {\em Physical review letters}, vol.~79, no.~2, p.~325, 1997.

\bibitem{Nielsen-Chuang}
M.~A. Nielsen and I.~Chuang, {\em Quantum computation and quantum information}.
\newblock Cambridge, Cambridge University Press, 2000.

\bibitem{GMD}
A.~Galindo and M.~A. Martin-Delgado, ``Information and computation: Classical
  and quantum aspects,'' {\em Reviews of Modern Physics}, vol.~74, no.~2,
  p.~347, 2002.

\bibitem{Brandao-Svore}
F.~G. Brandao and K.~M. Svore, ``Quantum speed-ups for solving semidefinite
  programs,'' in {\em Foundations of Computer Science (FOCS), 2017 IEEE 58th
  Annual Symposium on}, pp.~415--426, IEEE, 2017.

\bibitem{QuantumSDP1}
F.~G. Brandao, A.~Kalev, T.~Li, C.~Y.-Y. Lin, K.~M. Svore, and X.~Wu,
  ``Exponential quantum speed-ups for semidefinite programming with
  applications to quantum learning,'' {\em arXiv preprint arXiv:1710.02581},
  2017.

\bibitem{QuantumSDP2}
J.~Van~Apeldoorn, A.~Gily{\'e}n, S.~Gribling, and R.~de~Wolf, ``Quantum
  sdp-solvers: Better upper and lower bounds,'' in {\em Foundations of Computer
  Science (FOCS), 2017 IEEE 58th Annual Symposium on}, pp.~403--414, IEEE,
  2017.

\bibitem{QuantumSDP3}
J.~van Apeldoorn and A.~Gily{\'e}n, ``Improvements in quantum sdp-solving with
  applications,'' {\em arXiv preprint arXiv:1804.05058}, 2018.

\bibitem{QuantumSDP4}
S.~Chakrabarti, A.~M. Childs, T.~Li, and X.~Wu, ``Quantum algorithms and lower
  bounds for convex optimization,'' {\em arXiv preprint arXiv:1809.01731},
  2018.

\bibitem{Khachiyan}
L.~G. Khachiyan, ``A polynomial algorithm in linear programming,'' in {\em
  Doklady Academii Nauk SSSR}, vol.~244, pp.~1093--1096, 1979.

\bibitem{Karmarkar}
N.~Karmarkar, ``A new polynomial-time algorithm for linear programming,'' in
  {\em Proceedings of the sixteenth annual ACM symposium on Theory of
  computing}, pp.~302--311, ACM, 1984.

\bibitem{Potra-Wright}
F.~A. Potra and S.~J. Wright, ``Interior-point methods,'' {\em Journal of
  Computational and Applied Mathematics}, vol.~124, no.~1-2, pp.~281--302,
  2000.

\bibitem{Predictor-CorrectorI}
S.~Mizuno, M.~J. Todd, and Y.~Ye, ``On adaptive-step primal-dual interior-point
  algorithms for linear programming,'' {\em Mathematics of Operations
  research}, vol.~18, no.~4, pp.~964--981, 1993.

\bibitem{conjugate_gradient}
J.~R. Shewchuk {\em et~al.}, ``An introduction to the conjugate gradient method
  without the agonizing pain,'' 1994.

\bibitem{numerical_recipies}
W.~H. Press, B.~P. Flannery, S.~A. Teukolsky, W.~T. Vetterling, {\em et~al.},
  {\em Numerical recipes}, vol.~2.
\newblock Cambridge university press Cambridge, 1989.

\bibitem{dense_exact}
V.~Pan and J.~Reif, ``Fast and efficient parallel solution of dense linear
  systems,'' {\em Computers \& Mathematics with Applications}, vol.~17, no.~11,
  pp.~1481--1491, 1989.

\bibitem{kerenidis2018quantum}
I.~Kerenidis and A.~Prakash, ``A quantum interior point method for lps and
  sdps,'' {\em arXiv preprint arXiv:1808.09266}, 2018.

\bibitem{nannicini2019fast}
G.~Nannicini, ``Fast quantum subroutines for the simplex method,'' {\em arXiv
  preprint arXiv:1910.10649}, 2019.

\bibitem{Average_performance}
K.~M. Anstreicher, J.~Ji, F.~A. Potra, and Y.~Ye, ``Average performance of a
  self--dual interior point algorithm for linear programming,'' in {\em
  Complexity in numerical optimization}, pp.~1--15, World Scientific, 1993.

\bibitem{Aspuru}
A.~Aspuru-Guzik, A.~D. Dutoi, P.~J. Love, and M.~Head-Gordon, ``Simulated
  quantum computation of molecular energies,'' {\em Science}, vol.~309,
  no.~5741, pp.~1704--1707, 2005.

\bibitem{Mezzacapo}
A.~Kandala, A.~Mezzacapo, K.~Temme, M.~Takita, M.~Brink, J.~M. Chow, and J.~M.
  Gambetta, ``Hardware-efficient variational quantum eigensolver for small
  molecules and quantum magnets,'' {\em Nature}, vol.~549, no.~7671, p.~242,
  2017.

\bibitem{Solano}
M.-H. Yung, J.~Casanova, A.~Mezzacapo, J.~Mcclean, L.~Lamata, A.~Aspuru-Guzik,
  and E.~Solano, ``From transistor to trapped-ion computers for quantum
  chemistry,'' {\em Scientific reports}, vol.~4, p.~3589, 2014.

\bibitem{Innsbruck}
C.~Hempel, C.~Maier, J.~Romero, J.~McClean, T.~Monz, H.~Shen, P.~Jurcevic,
  B.~Lanyon, P.~Love, R.~Babbush, {\em et~al.}, ``Quantum chemistry
  calculations on a trapped-ion quantum simulator,'' {\em arXiv preprint
  arXiv:1803.10238}, 2018.

\bibitem{ReviewQCh}
Y.~Cao, J.~Romero, J.~P. Olson, M.~Degroote, P.~D. Johnson, M.~Kieferov{\'a},
  I.~D. Kivlichan, T.~Menke, B.~Peropadre, N.~P. Sawaya, {\em et~al.},
  ``Quantum chemistry in the age of quantum computing,'' {\em arXiv preprint
  arXiv:1812.09976}, 2018.

\bibitem{HHL}
A.~W. Harrow, A.~Hassidim, and S.~Lloyd, ``Quantum algorithm for linear systems
  of equations,'' {\em Physical review letters}, vol.~103, no.~15, p.~150502,
  2009.

\bibitem{DenseHHL}
L.~Wossnig, Z.~Zhao, and A.~Prakash, ``Quantum linear system algorithm for
  dense matrices,'' {\em Physical review letters}, vol.~120, no.~5, p.~050502,
  2018.

\bibitem{coppersmith1990matrix}
D.~Coppersmith and S.~Winograd, ``Matrix multiplication via arithmetic
  progressions,'' {\em Journal of symbolic computation}, vol.~9, no.~3,
  pp.~251--280, 1990.

\bibitem{Arora-Kale}
S.~Arora and S.~Kale, ``A combinatorial, primal-dual approach to semidefinite
  programs,'' in {\em Proceedings of the thirty-ninth annual ACM symposium on
  Theory of computing}, pp.~227--236, ACM, 2007.

\bibitem{kerenidis2017recommendation}
I.~Kerenidis and A.~Prakash, ``Quantum recommendation systems,'' in {\em
  Proceedings of the 8th Innovations in Theoretical Computer Science
  Conference}, 2017.

\bibitem{FEM}
A.~Montanaro and S.~Pallister, ``Quantum algorithms and the finite element
  method,'' {\em Physical Review A}, vol.~93, no.~3, p.~032324, 2016.

\bibitem{qubitization}
G.~H. Low and I.~L. Chuang, ``Hamiltonian simulation by qubitization,'' {\em
  Quantum}, vol.~3, p.~163, 2019.

\bibitem{Dense_hamiltonian_simulation}
C.~Wang and L.~Wossnig, ``A quantum algorithm for simulating non-sparse
  hamiltonians,'' {\em arXiv preprint arXiv:1803.08273}, 2018.

\bibitem{QLSAchilds}
A.~M. Childs, R.~Kothari, and R.~D. Somma, ``Quantum algorithm for systems of
  linear equations with exponentially improved dependence on precision,'' {\em
  SIAM Journal on Computing}, vol.~46, no.~6, pp.~1920--1950, 2017.

\bibitem{SPAI-QLSA}
B.~D. Clader, B.~C. Jacobs, and C.~R. Sprouse, ``Preconditioned quantum linear
  system algorithm,'' {\em Physical review letters}, vol.~110, no.~25,
  p.~250504, 2013.

\bibitem{State_prep_1}
V.~V. Shende, S.~S. Bullock, and I.~L. Markov, ``Synthesis of quantum-logic
  circuits,'' {\em IEEE Transactions on Computer-Aided Design of Integrated
  Circuits and Systems}, vol.~25, no.~6, pp.~1000--1010, 2006.

\bibitem{State_prep_grover}
L.~Grover and T.~Rudolph, ``Creating superpositions that correspond to
  efficiently integrable probability distributions,'' {\em arXiv preprint
  quant-ph/0208112}, 2002.

\bibitem{State_prep_2}
Y.~R. Sanders, G.~H. Low, A.~Scherer, and D.~W. Berry, ``Black-box quantum
  state preparation without arithmetic,'' {\em Physical review letters},
  vol.~122, no.~2, p.~020502, 2019.

\bibitem{block-encoding_2}
A.~Gily{\'e}n, Y.~Su, G.~H. Low, and N.~Wiebe, ``Quantum singular value
  transformation and beyond: exponential improvements for quantum matrix
  arithmetics,'' {\em arXiv preprint arXiv:1806.01838}, 2018.

\bibitem{Ambainis}
A.~Ambainis, ``Variable time amplitude amplification and quantum algorithms for
  linear algebra problems,'' in {\em STACS'12 (29th Symposium on Theoretical
  Aspects of Computer Science)}, vol.~14, pp.~636--647, LIPIcs, 2012.

\bibitem{ye_termination}
Y.~Ye, ``On the finite convergence of interior-point algorithms for linear
  programming,'' {\em Mathematical Programming}, vol.~57, no.~1-3,
  pp.~325--335, 1992.

\bibitem{GroverMin}
L.~A.~B. Kowada, C.~Lavor, R.~Portugal, and C.~M. De~Figueiredo, ``A new
  quantum algorithm for solving the minimum searching problem,'' {\em
  International Journal of Quantum Information}, vol.~6, no.~03, pp.~427--436,
  2008.

\bibitem{tang2018quantum}
E.~Tang, ``A quantum-inspired classical algorithm for recommendation systems,''
  {\em arXiv preprint arXiv:1807.04271}, 2018.

\bibitem{chia2018quantum}
N.-H. Chia, H.-H. Lin, and C.~Wang, ``Quantum-inspired sublinear classical
  algorithms for solving low-rank linear systems,'' {\em arXiv preprint
  arXiv:1811.04852}, 2018.

\bibitem{arrazola2019quantum}
J.~M. Arrazola, A.~Delgado, B.~R. Bardhan, and S.~Lloyd, ``Quantum-inspired
  algorithms in practice,'' {\em arXiv preprint arXiv:1905.10415}, 2019.

\bibitem{nagaj2009fast}
D.~Nagaj, P.~Wocjan, and Y.~Zhang, ``Fast amplification of qma,'' {\em arXiv
  preprint arXiv:0904.1549}, 2009.

\end{thebibliography}



\appendix

\section{\label{sec:epsilon'} Calculation of the error $\epsilon'$ in the Corrector step}

In this appendix we want to calculate the size of the error $\epsilon'$ we need in order to make the Corrector step successfully output a point within $\mathcal{N}(1/4)$, and its comparison with the complexity of the quantum subroutine.

To prove this suppose we define $\bm{x}$ as the concatenation of $\bm{x}^t$ and $\tau^t$, and $\bm{s}$ as concatenating $\bm{s}^t$ and $k^t$. Call $\bm{x}_0$ and $\bm{s}_0$ the exact arithmetic solution, so that being $\epsilon'$ the error of the quantum subroutine
\begin{equation}
    \bm{x}=\bm{x}_0+\epsilon' \bm{x}_1; \quad \bm{s}=\bm{s}_0+\epsilon \bm{s}_1; \quad ||\bm{s}_1||=||\bm{x}_1||\leq n+1. \label{x1 and s1}
\end{equation}
since each entry will have an error of $\epsilon'$ at most, and then each entry in $\bm{x}_1$ and $\bm{s}_1$ are, say, at most $1$.
Using \eqref{central path neighbourhood}, we can see that
\begin{equation}
    \left|\left|X_0\bm{s}_0-\mathbf{1}\frac{\bm{x}^T_0\bm{s}_0}{n+1}\right|\right|\leq \frac{1}{4\sqrt{2}}\frac{\bm{x}^T_0\bm{s}_0}{n+1}, \label{hypothesis failure}
\end{equation}
and we want to calculate how small $\epsilon'$ needs to be in order to comply with  
\begin{equation}
    \left|\left|X\bm{s}-\mathbf{1}\frac{\bm{x}^T\bm{s}}{n+1}\right|\right|\leq \frac{1}{4}\frac{\bm{x}^T\bm{s}}{n+1}. \label{target failure}
\end{equation}
Expanding, to leading order $O(\epsilon')$
\begin{align}
    &\left|\left|X\bm{s}-\mathbf{1}\frac{\bm{x}^T\bm{s}}{n+1}\right|\right|=\\
    &\left|\left|X_0\bm{s}_0+\epsilon'(X_0\bm{s}_1+X_1\bm{s}_0)\right.\right.\\
    &\left.\left.-\mathbf{1}\frac{\bm{x}_0^T\bm{s}_0+\epsilon'(\bm{x}_0^T\bm{s}_1+\bm{x}_1^T\bm{s}_0)}{n+1}\right|\right|\\
    &\leq  \left|\left|X_0\bm{s}_0-\mathbf{1}\frac{\bm{x}^T_0\bm{s}_0}{n+1}\right|\right|\\
    &+ \epsilon' \left|\left|X_0\bm{s}_1-\mathbf{1}\frac{\bm{x}^T_0\bm{s}_1}{n+1}\right|\right|\\
    &+ \epsilon'  \left|\left|X_1\bm{s}_0-\mathbf{1}\frac{\bm{x}^T_1\bm{s}_0}{n+1}\right|\right|.
\end{align}
The first term is clearly our hypothesis \eqref{hypothesis failure}, so let us calculate one of other two terms (calculating one is the same as calculating the other, they are symmetrical). Let $\bm{x}_i = (\bar{x}_{i,1},...,\bar{x}_{i,n+1})^T$ for $i\in\{0,1\}$ and similarly for $\bm{s}_i$, and let us expand the expression.
\begin{align}
  &\left|\left|X_1\bm{s}_0-\mathbf{1}\frac{\bm{x}^T_1\bm{s}_0}{n+1}\right|\right|=\\
  &\sqrt{\sum_{i=1}^{n+1}\left(\bar{x}_{1,i}\bar{s}_{0,i}-\frac{1}{n+1}\sum_{j=1}^{n+1}\bar{x}_{1,j}\bar{s}_{0,j}\right)^2}=\\
  &\left[\sum_{i=1}^{n+1}\left(\bar{x}^2_{1,i}\bar{s}^2_{0,i}-\frac{2}{n+1}\bar{x}_{1,i}\bar{s}_{0,i}\sum_{j=1}^{n+1}\bar{x}_{1,j}\bar{s}_{0,j}\right.\right.\\
  &+\left.\left.\frac{1}{(n+1)^2}\left(\sum_{j=1}^{n+1}\bar{x}_{1,j}\bar{s}_{0,j}\right)^2 \right) \right]^{1/2}=\\
  &\left[\sum_{i=1}^{n+1}\bar{x}^2_{1,i}\bar{s}^2_{0,i}-\frac{2}{n+1}\sum_{i,j=1}^{n+1}\bar{x}_{1,i}\bar{s}_{0,i}\bar{x}_{1,j}\bar{s}_{0,j}\right.\\  
  &+\left.\frac{n+1}{(n+1)^2}\left( \sum_{i,j=1}^{n+1}\bar{x}_{1,i}\bar{s}_{0,i}\bar{x}_{1,j}\bar{s}_{0,j} \right)\right]^{1/2}
\end{align}
Now we can see that the second term partially cancels out with the third
\begin{align}
  &\left[\sum_{i=1}^{n+1}\bar{x}^2_{1,i}\bar{s}^2_{0,i}-\frac{1}{n+1}\sum_{i,j=1}^{n+1}\bar{x}_{1,i}\bar{s}_{0,i}\bar{x}_{1,j}\bar{s}_{0,j}\right]^{1/2}  \label{simplified norm}\\
  &=\left[||X_1\bm{s}_0||^2-\frac{1}{n+1}(\bm{x}_1^T\bm{s}_0)^2\right]^{1/2}
\end{align}
Therefore, we can write
\begin{align}
    &\left|\left|Xs-\mathbf{1}\frac{\bm{x}^Ts}{n+1}\right|\right|\leq \frac{1}{4\sqrt{2}}\frac{\bm{x}^T_0\bm{s}_0}{n+1} \label{eq 50} \\
    &+ \epsilon'\left(\left[||X_1\bm{s}_0||^2-\frac{1}{n+1}(\bm{x}_1^T\bm{s}_0)^2\right]^{1/2}\right.\label{eq 51}\\
    &+\left.\left[||X_0\bm{s}_1||^2-\frac{1}{n+1}(\bm{x}_0^T\bm{s}_1)^2\right]^{1/2}\right)\label{eq 52}
\end{align}
Enforcing \eqref{target failure} can be done if we choose $\epsilon'$ such that
\begin{align}
    &\frac{1}{4\sqrt{2}}\frac{\bm{x}^T_0\bm{s}_0}{n+1} + \epsilon'\left(\left[||X_1\bm{s}_0||^2-\frac{1}{n+1}(\bm{x}_1^T\bm{s}_0)^2\right]^{1/2}\right.\\
    &+\left.\left[||X_0\bm{s}_1||^2-\frac{1}{n+1}(\bm{x}_0^T\bm{s}_1)^2\right]^{1/2}\right)\\
    &\leq \frac{1}{4}\frac{\bm{x}^T \bm{s}}{n+1}
     = \frac{1}{4}\frac{(\bm{x}_0+\epsilon' \bm{x}_1)^T(\bm{s}_0+\epsilon' \bm{s}_1)}{n+1}.
\end{align}

To leading order $O(\epsilon')$ that means
\begin{align}
    &\frac{1}{4\sqrt{2}}\frac{\bm{x}^T_0\bm{s}_0}{n+1} + \epsilon'\left(\left[||X_1\bm{s}_0||^2-\frac{1}{n+1}(\bm{x}_1^T\bm{s}_0)^2\right]^{1/2}\right.\\
    &+\left.\left[||X_0\bm{s}_1||^2-\frac{1}{n+1}(\bm{x}_0^T\bm{s}_1)^2\right]^{1/2}\right)\leq\\
    &\frac{1}{4}\frac{\bm{x}_0^T\bm{s}_0}{n+1}+\frac{\epsilon'}{4}\frac{\bm{x}_1^T\bm{s}_0+\bm{x}_0^T\bm{s}_1}{n+1},
\end{align}
or equivalently
\begin{align}
    &\frac{2-\sqrt{2}}{8}\frac{\bm{x}^T_0\bm{s}_0}{n+1}\geq\\
    &\epsilon'\left(\left[||X_1\bm{s}_0||^2-\frac{1}{n+1}(\bm{x}_1^T\bm{s}_0)^2\right]^{1/2}\right.\\
    &+\left.\left[||X_0\bm{s}_1||^2-\frac{1}{n+1}(\bm{x}_0^T\bm{s}_1)^2\right]^{1/2}-\frac{1}{4}\frac{\bm{x}_1^T\bm{s}_0+\bm{x}_0^T\bm{s}_1}{n+1}\right),
\end{align}
implying
\begin{widetext}
\begin{equation}
    \epsilon'\leq \frac{\frac{2-\sqrt{2}}{8}\frac{\bm{x}^T_0\bm{s}_0}{n+1}}{\left[||X_1\bm{s}_0||^2-\frac{1}{n+1}(\bm{x}_1^T\bm{s}_0)^2\right]^{1/2}
    +\left[||X_0\bm{s}_1||^2-\frac{1}{n+1}(\bm{x}_0^T\bm{s}_1)^2\right]^{1/2}-\frac{1}{4}\frac{\bm{x}_1^T\bm{s}_0+\bm{x}_0^T\bm{s}_1}{n+1}}. \label{epsilon'}
\end{equation}
\end{widetext}
Let us start analyzing the denominator. The worst case is that the denominator is very large so it forces the error to be small. The fraction $\frac{1}{4}\frac{\bm{x}_1^T\bm{s}_0+\bm{x}_0^T\bm{s}_1}{n+1}$ could turn negative, thus increasing the denominator. But we can see that its influence will be low due to its denominator $n+1$. Due to \eqref{x1 and s1}, we can expect $\bm{x}_1^T\bm{s}_0\leq (n+1)\max_i{|\bar{s}_{0,i}|} = O(n)$ and $\bm{x}_0^T\bm{s}_1\leq (n+1)\max_i{|\bar{x}_{0,i}|}=O(n)$. Therefore $\frac{1}{4}\frac{\bm{x}_1^T\bm{s}_0+\bm{x}_0^T\bm{s}_1}{n+1} = O(1)$.

Additionally, if all entries in $\bm{x}_0$ and $\bm{s}_0$ are $O(1)$ it is easy to check that $\left|\left|X_0\bm{s}_1\right|\right|^2  = O(n)$ and $\left|\left|X_1\bm{s}_0\right|\right|^2 = O(n)$. Therefore the denominator will be $O(n^{1/2})$.

The numerator of \eqref{epsilon'} is a bit more tricky. In this case the worst case is when it is small. What seems the biggest problem is that \cite{Predictor-Corrector} tells us that in the exact solution $\bm{x}_0^T\bm{s}_0 = 0$, and furthermore it is divided by $n+1$. Therefore we can see that in the exact solution $\epsilon$ should be $0$ (which is what we should expect). What matters is the convergence speed. At the beginning recall that $\bm{x}_0 = \bm{s}_0 = [1,...,1]^T$. Therefore, at the very beginning $\frac{\bm{x}_0^T\bm{s}_0}{n+1}=1$. According to \cite{Predictor-Corrector}, every two iterations $\bm{x}^T_0\bm{s}_0$ should decrease at a rate $(1-1/(\sqrt[4]{8}\sqrt{n+1}))$. The question therefore is what happens after $O(L\sqrt{n}\bar{t})$ iterations (the iterations of the algorithm).

This finally gives us the threshold for $\epsilon'$ we would like in order to stay inside $\mathcal{N}(1/4)$:
\begin{equation}
\begin{split}
    \epsilon'&\leq \lim_{n\rightarrow\infty}O(n^{-1/2})\left(1-\frac{1}{\sqrt[4]{8}\sqrt{n}}\right)^{O(L\sqrt{n}\bar{t})} \\
    &= O(n^{-1/2})e^{-O(L\bar{t})}. \label{epsilon' N(1/4)}
\end{split}
\end{equation}
Since $L=\log n$ in the usual case and $\bar{t}=\max[\log((\bm{x}^0)^T( \bm{s}^0)/(\epsilon_1\epsilon_3^2)), \log(||(\bar{\bm{b}}^T,\bar{\bm{c}}^T)||/\epsilon_2\epsilon_3)]$, the error threshold is $\epsilon' = O(n^{-\text{poly}\log n-1/2})$.

On the other hand, the complexity on the precision of the algorithm is $O(\epsilon^{-2})$ and the difference in complexity in $n$ that we stated in tabla \ref{tabla comparativa complejidad} between our algorithm and the `best classical' algorithm is $O(\sqrt{n})$. This means that in order to maintain the quantum advantage one would want $\epsilon\leq O(n^{-1/4})$.
Therefore we have seen that setting an $\epsilon'$ small enough might be too expensive computationally in general for the quantum subroutine.

\section{\label{sec:epsilon''} Gradient descent for shifting the output of the corrector step.}

We have seen that setting an $\epsilon'$ small enough might be too expensive computationally. Therefore let us try something: choose a small precision on the error for the corrector step $\epsilon' = O(n^{-1/a})$, and then, once we get the result, perform one step gradient descent of the size $\epsilon''$ towards $\mathcal{N}(1/4)$.

For that call, using the definition of $\mathcal{N}(\beta)$ \eqref{central path neighbourhood},
\begin{equation}
    g(x,s)=\left|\left|X\bm{s}-\mathbf{1}\frac{\bm{x}^T\bm{s}}{n+1}\right|\right|^2-\beta^{2}\left(\frac{\bm{x}^T\bm{s}}{n+1}\right)^2.
\end{equation}
Using the equivalent to \eqref{simplified norm}, we can rewrite it like
\begin{align}
     &g(x,s)=||Xs-\mathbf{1}\mu||^2-\beta^2\mu^2\\
     &=\left(\sum_i \bar{x}_i^2 \bar{s}_i^2-\frac{1}{n+1}\sum_{i,j=1}^{n+1}\bar{x}_{i}\bar{s}_{i}\bar{x}_{j}\bar{s}_{j}\right)\\
     &- \frac{\beta^2}{(n+1)^2}\left( \sum_{i,j=1}^{n+1}\bar{x}_i \bar{s}_i\bar{x}_j \bar{s}_j\right)   \\
     &=\sum_i \bar{x}_i^2 \bar{s}_i^2-\left(\frac{\beta^2+(n+1)}{(n+1)^2}\right)\sum_{i, j=1}^{n+1}\bar{x}_i \bar{s}_i\bar{x}_j \bar{s}_j. \label{g definition}
\end{align}
Notice that the points that are in $\mathcal{N}(\beta)$ are those for which $g(x,s)\leq 0$. We calculate the gradient, calling $B=\frac{\beta^2+(n+1)}{(n+1)^2}=O(n^{-1})$:
\begin{equation}
    \frac{d g(x,s)}{d \bar{x}_k} = 2\bar{x}_k \bar{s}_k^2-2B\sum_{i}\bar{x}_i \bar{s}_i \bar{s}_k. \label{dg dx}
\end{equation}
Clearly,
\begin{equation}
    \frac{d g(x,s)}{d \bar{s}_k} = 2\bar{s}_k \bar{x}_k^2-2B\sum_{i}\bar{s}_i \bar{x}_i \bar{x}_k.  \label{dg ds}
\end{equation}
The idea is now, supposing that
\begin{equation}
    \left|\left|X_0\bm{s}_0-\mathbf{1}\frac{\bm{x}_0^T\bm{s}_0}{n+1}\right|\right|^2\leq\beta^{2}\left(\frac{\bm{x}^T_0\bm{s}_0}{n+1}\right)^2,
\end{equation}
if we can find $\epsilon'' = O( n^{-1/a})$ such that
\begin{align} \label{result seeked}
    &\left|\left|X\bm{s}-\epsilon'' \left(X\bm{\frac{dg}{ds}}+S\bm{\frac{dg}{dx}}\right)\right.\right.\\
    &\left.\left.-\mathbf{1}\frac{1}{n+1}\left(\bm{x}^T\bm{s}-\epsilon''\left( \bm{x}^T\bm{\frac{dg}{ds}}+ \bm{s}^T\bm{\frac{dg}{dx}}\right)\right)\right|\right|^2\\
    &\leq\frac{\beta^2}{(n+1)^2}\left(\bm{x}^T\bm{s}-\epsilon''\left( \bm{x}^T\bm{\frac{dg}{ds}}+ \bm{s}^T\bm{\frac{dg}{dx}}\right)\right)^2.
\end{align}
For that, as it will be very useful, first calculate to leading order $O(\epsilon'')$
\begin{equation}
\begin{split}
    &\sum_i \left(\bar{x}_i-\epsilon''\frac{dg}{d\bar{x}_i}\right)\left(\bar{s}_i-\epsilon''\frac{dg}{d\bar{s}_i}\right)=\\
    & \sum_i\left[ \bar{x}_i\bar{s}_i-2\epsilon''(\bar{s}_i^2+\bar{x}_i^2)\left(\bar{s}_i\bar{x}_i-B\sum_j \bar{x}_j\bar{s}_j\right)\right]. \label{expansion epsilon''}
\end{split}
\end{equation}
We can write, to leading order $O(\epsilon)$, how much is $||Xs||^2$ in the new point:
\begin{equation}
\begin{split}
    &\sum_i \left(\bar{s}_i-\epsilon''\frac{dg}{d\bar{s}_i}\right)^2\left(\bar{x}_i-\epsilon''\frac{dg}{d\bar{x}_i}\right)^2 =\\
    &\sum_i \left[\bar{x}_i^2\bar{s}_i^2-4\epsilon''\bar{x}_i\bar{s}_i(\bar{s}_i^2+\bar{x}_i^2)\left(\bar{s}_i\bar{x}_i-B\sum_j \bar{x}_j\bar{s}_j\right)\right].
    \label{expansion epsilon'' |Xs|}
\end{split}
\end{equation}

The next thing we want to calculate is, to leading order $O(\epsilon)$, the new value of $(\bm{x}^T\bm{s})^2$
\begin{equation}
\begin{split}
    &\sum_{i,j}\left(\bar{s}_i-\epsilon''\frac{dg}{d\bar{s}_i}\right)\left(\bar{x}_i-\epsilon''\frac{dg}{d\bar{x}_i}\right)\\
    &\left(\bar{s}_j-\epsilon''\frac{dg}{d\bar{s}_j}\right)\left(\bar{x}_j-\epsilon''\frac{dg}{d\bar{x}_j}\right)\\
    &= \sum_{i,j}\bar{x}_i\bar{x}_j\bar{s}_i\bar{s}_j\\
    &-\sum_{i,j}4\epsilon''\bar{x}_j\bar{s}_j(\bar{s}_i^2+\bar{x}_i^2)\left(\bar{s}_i\bar{x}_i-B\sum_k \bar{x}_k\bar{s}_k\right)
    \label{expansion epsilon'' (xs)}
\end{split}
\end{equation}

We now have all the parts we need to calculate the result we were seeking.
Let us try to check that we can make 
\begin{equation}
\begin{split}
    g\left(\bm{x}-\epsilon''\bm{\frac{dg}{dx}}, \bm{s}-\epsilon''\bm{\frac{dg}{ds}}\right)\leq 0.
\end{split}
\end{equation}
Let's check it

\begin{equation}
\begin{split}
    &g\left(\bm{x}-\epsilon''\bm{\frac{dg}{dx}}, \bm{s}-\epsilon''\bm{\frac{dg}{ds}}\right)=\\
    &=\sum_i \bar{x}_i^2 \bar{s}_i^2-B\sum_{i, j=1}^{n+1}\bar{x}_i \bar{s}_i\bar{x}_j \bar{s}_j+\\
    &-4\epsilon''\left[ \sum_i \bar{x}_i\bar{s}_i(\bar{s}_i^2+\bar{x}_i^2)\left(\bar{s}_i\bar{x}_i-B\sum_j \bar{x}_j\bar{s}_j\right)\right.\\
    &\left.-B\sum_{i,j}\bar{x}_j\bar{s}_j(\bar{s}_i^2+\bar{x}_i^2)\left(\bar{s}_i\bar{x}_i-B\sum_k \bar{x}_k\bar{s}_k\right)\right]
    \label{calculation of epsilon''}
\end{split}
\end{equation}
This allows us to calculate the approximate $\epsilon''$ in time $O(n)$ such that the point is in $\mathcal{N}(1/4)$. Is this $\epsilon''$ large? To answer that question expand in $\bm{x} = \bm{x}_0 + \epsilon' \bm{x}_1$ and $\bm{s} = \bm{s}_0 + \epsilon' \bm{s}_1$
\begin{equation}
\begin{split}
    &g\left(\bm{x}-\epsilon''\bm{\frac{dg}{dx}}, \bm{s}-\epsilon''\bm{\frac{dg}{ds}}\right)=\\
    &=\sum_i \bar{x}_{i,0}^2 \bar{s}_{i,0}^2-B\sum_{i, j=1}^{n+1}\bar{x}_{i,0} \bar{s}_{i,0}\bar{x}_{j,0} \bar{s}_{j,0}+\\
    &+2\epsilon'\left[\sum_i ( \bar{x}_{i,0}\bar{x}_{i,1} \bar{s}_{i,0}^2+\sum_i \bar{x}_{i,0}^2 \bar{s}_{i,0}\bar{s}_{i,1}) \right.\\
    &\left.-B\sum_{i,j}(\bar{x}_{i,1} \bar{s}_{i,0}\bar{x}_{j,0} \bar{s}_{j,0}+\bar{x}_{i,0} \bar{s}_{i,1}\bar{x}_{j,0} \bar{s}_{j,0})\right]\\
    &-4\epsilon''\left[ \sum_i \bar{x}_i\bar{s}_i(\bar{s}_i^2+\bar{x}_i^2)\left(\bar{s}_i\bar{x}_i-B\sum_j \bar{x}_j\bar{s}_j\right)\right.\\
    &\left.-B\sum_{i,j}\bar{x}_j\bar{s}_j(\bar{s}_i^2+\bar{x}_i^2)\left(\bar{s}_i\bar{x}_i-B\sum_k \bar{x}_k\bar{s}_k\right)\right].
    \label{epsilon'' is O(epsilon')}
\end{split}
\end{equation}
We know by hypothesis that the second line of \eqref{epsilon'' is O(epsilon')} is less than 0 (the exact point is in $\mathcal{N}(1/4)$). Therefore we want the term of $\epsilon''$ to cancel that of $\epsilon'$. Notice that this means that $\epsilon'' = O(\epsilon')$, since both terms in the square brackets are of size $O(n)$. Therefore, we are only introducing an error of size $O(\epsilon')$, which is what we already had, but we can now ensure that the output is in $\mathcal{N}(1/4)$. \eqref{calculation of epsilon''} explains how to calculate $\epsilon''$ such that, to order $O(\epsilon'')$ we are inside $\mathcal{N}(1/4)$; and \eqref{epsilon'' is O(epsilon')} certifies that $\epsilon'' = O(\epsilon')$.

Since in the expansion we have only consider terms up to $\epsilon''$, there is the possibility that after this shift we are still outside $\mathcal{N}(1/4)$. However, in this case, we will only be $\epsilon''^2 = O(\epsilon'^2)$ away from $\mathcal{N}(1/4)$, so iterating this process we can make this error $O(\epsilon''^{2^i})$, which decreases quickly on the iteration $i$. Alternatively one could take into account further terms with higher exponents of $\epsilon''$ in the expansions \eqref{expansion epsilon''}, \eqref{expansion epsilon'' |Xs|} and \eqref{expansion epsilon'' (xs)}.

Further, one could also wonder if the gradient step could converge towards different local minima of $g(x,s)$ than those in the central path. However, it is easy to check using \eqref{dg dx} and \eqref{dg ds} that all local minima (or maxima or saddle points) fulfil either $\bm{x}_i=0$ and $\bm{s}_i=0$, or
\begin{equation}
    x_i s_i -B\sum_j x_j s_j = 0
\end{equation}
for all $i\in \{1,...,n+1\}$. Substituting those points in the definition of $g(x,s)$ \eqref{g definition}, 
\begin{equation}
\begin{split}
    g &= \sum_{i=1}^{n+1} x_i^2 s_i^2-\left(\frac{\beta^2+(n+1)}{(n+1)^2}\right)(n+1)\sum_{i=1}^{n+1}x_i^2 s_i^2\\
    &= \left(1-\frac{\beta^2}{n+1}-1\right)\sum_{i=1}^{n+1} x_i^2 s_i^2 \leq 0.
\end{split}
\end{equation}
This means that all those points are within $\mathcal{N}(\beta)$, so the gradient step will be taken towards $\mathcal{N}(\beta)$.

With this we conclude that with an $O(\epsilon')$-shift to the output of the already $O(\epsilon')$-precise Corrector step we can ensure that the point is in $\mathcal{N}(1/4)$. 

\section{Convergence of the algorithm} \label{sec:convergence}
The legitimacy of the previous procedure rests on the fact that we only have to check that we do not get out of the neighbourhood of the central path. However, a question remains: does the algorithm still converges, with this modification, in $O(n^{1/2}L)$ steps? In this appendix we analyse this question and find a positive answer. The references we will be using are \cite{Predictor-CorrectorI}, mainly, and \cite{Predictor-Corrector}.

Let us first review how the convergence is studied in the original article \cite{Predictor-Corrector}. The convergence happens when the duality gap $\mu$ closes. In those references, one proves that in the corrector step $\mu^k = \mu^{k+1}$. The convergence happens in the predictor corrector, where the duality gap $\mu^{k+1} = (1-\alpha)\mu^k$. Lets see these two things.

For convenience define $\bm{x}^t$ concatenating $\bm{x}^t$ and $\tau^t$; $\bm{s}^t$ as the concatenation of $\bm{s}^t$ and $\kappa^t$; and similarly with $\bm{d_x}$, $d_\tau$, and $\bm{d_s}$ and $d_\kappa$. The starting equation is 
\begin{equation}
\begin{split}
    \mu^{t+1}(\delta) &= \frac{(\bm{x}^{t+1})^T \bm{s}^{t+1}}{n+1} =\\
    &\frac{(\bm{x}^{t})^T \bm{s}^{t} + \delta ((\bm{x}^t)^T\bm{d_s} +(\bm{s}^t)^T \bm{d_x}) +\delta^2 \bm{d_x}^T \bm{d_s} }{n+1}.
    \label{mu delta corrector}
\end{split}
\end{equation}
Since the exact solution fulfils equation \eqref{eq12}, that would mean that 
\begin{equation}
   X^t \bm{d_s} + S^t \bm{d_x} = \gamma^t \mu^t \bm{1} - X^t\bm{s}^t.
\end{equation}
In practice, since we get an approximation solution, we can substitute the term multiplied by $\delta$ in \eqref{mu delta corrector} with $\gamma^t\mu^t (n+1)-(\bm{x}^t)^T\bm{s}^t + (\bm{x}^t)^T\bm{\epsilon_x} + (\bm{s}^t)^T\bm{\epsilon_s}$, where $\bm{\epsilon_x}$ and $\bm{\epsilon_s}$ are the error vectors in the estimation of $\bm{d_x}$ and $\bm{d_s}$ respectively. 

On the other hand, using equation \eqref{eq11} for the exact solution, and the second part of theorem 5 of \cite{Predictor-Corrector}, we know that $(\bm{d_x})^T \bm{d_s} = 0$. In our case though, the equation will not be exact due to errors, and thus, the term multiplied by $\delta^2$ in \eqref{mu delta corrector} can be substituted with $\bm{d_x}^T \bm{\epsilon_s} + \bm{d_s}^T \bm{\epsilon_x}$. Therefore, in the approximate case, we can rewrite \eqref{mu delta corrector} for the corrector step ($\delta = 1 = \mu^t$) as
\begin{equation}
\begin{split}
    \mu^{t+1} &= \frac{(\bm{x}^{t})^T \bm{s}^{t} +  (-(\bm{x}^t)^T\bm{s}^t + (\bm{x}^t)^T\bm{\epsilon_x} + (\bm{s}^t)^T\bm{\epsilon_s})}{n+1}\\
    &+ \frac{(\bm{d_x}^T \bm{\epsilon_x} + \bm{d_s}^T \bm{\epsilon_x})}{n+1} + \mu^t.
    \label{mu delta corrector approximate}
\end{split}
\end{equation}
From this it follows that in the worst case
\begin{equation}
    \mu^{t+1} = \mu^t + O(\epsilon),
\end{equation}
where $\epsilon$ is the norm of each of the entries of $\bm{\epsilon_x}$ and $\bm{\epsilon_s}$, if the entries of $\bm{x}^t$, $\bm{s}^t$, $\bm{d_x}$ and $\bm{d_s}$ are $O(1)$.

The predictor step is a bit more involved, since $\delta$ is not fixed. The calculation, though, is similar, and we get
\begin{equation}
    \mu^{t+1}(\delta) = (1-\delta)\mu^{t} + (1 + \delta) O(\epsilon),
    \label{approximate mu predictor}
\end{equation}
where the term $(1-\delta)\mu^t$ is the exact value we would recover, and $O(\epsilon)$ the part due to the error.
Then question that remains is, how large can we make $\delta$ in the Predictor step? To answer this question, remember that condition for the Predictor step to end up in $\mathcal{N}(1/2)$ is
\begin{equation}
    ||X^{t+1}(\delta) \bm{s}^{t+1}(\delta)- \mu^{t+1}(\delta) 1|| \leq \frac{1}{2} \mu^{t+1}(\delta).
    \label{beta neighbourhood}
\end{equation}
We have already seen that the right hand side of \eqref{beta neighbourhood} is $O(\epsilon)$-approximate to the exact $\frac{1}{2} \mu^{t+1}$. The left-hand side on the other hand, is 
\begin{equation}
\begin{split}
    &||X^{t+1}(\delta) \bm{s}^{t+1}(\delta)- \mu^{t+1}(\delta) 1|| =\\
    & \left|\left| (X^t \bm{s}^t -\mu^t\bm{1}) \right. \right.\\
    & + \delta \left(X^t \bm{d_s} + S^t \bm{d_x} - \frac{(\bm{x}^t)^T \bm{d_s} + (\bm{s}^t)^T \bm{d_x}}{n+1} \mathbf{1} \right) \\
    & \left. \left.+ \delta^2 \left(D_x \bm{d_s} -\frac{\bm{d_x}^T \bm{d_s}}{n+1}\right)\right|\right|,
    \label{norm of Xs - mu}
\end{split}    
\end{equation}
where $D_x$ is the diagonal matrix corresponding to $\bm{d_x}$. In order to make this work, we will enforce a correction to make sure that \eqref{eq12} is fulfilled exactly, not only approximately. This means that we have to make $X^t \bm{d_x} + S^t \bm{d_s} = -X^t \bm{s}^t$. Let us expand it with the errors (recall that $\mu^t = 0$ for the predictor step).
\begin{equation}
    X^t(\bm{d_x} + \bm{\epsilon_x}) + S^t (\bm{d_s} + \bm{\epsilon_s}) = -X^t \bm{s}^t,
\end{equation}
The terms not containing either of the error vectors cancel out. To enforce that the previous equation is fulfilled exactly, we go one by one of the entries of $\bm{x}^t$ and $\bm{s}^t$ (complexity $O(n+1)$), and choose the largest value of both. If the modulus of the entry of $\bm{x}^t$ is larger, then substract $\epsilon'$ to ${\bm{d_x}}_i$ such that 
\begin{equation} \label{Predictor shift}
\begin{split}
    &\bm{x}_i^t({\bm{d_x}}_i + {\bm{\epsilon_x}}_i - \epsilon') + \bm{s}^t_i ({\bm{d_s}}_i + {\bm{\epsilon_s}}_i) = -\bm{x}_i^t \bm{s}_i^t, \iff\\
    & \bm{x}_i^t ( {\bm{\epsilon_x}}_i - \epsilon') + \bm{s}^t_i {\bm{\epsilon_s}}_i = 0 \iff \epsilon' =  \frac{\bm{s}_i^t}{\bm{x}_i^t}{\bm{\epsilon_s}}_i + {\bm{\epsilon_x}}_i .
\end{split}
\end{equation}
Clearly $\epsilon' = O(\epsilon)$, so the introduced error will be of the size of the original one. We do equivalently for the case when the modulus of the entry of $\bm{s}^t$ is larger than that of $\bm{x}^t$. If both entries are very small, then no correction is necessary. Then, we rewrite \eqref{norm of Xs - mu}
\begin{equation} \label{norm of Xs - mu delta}
\begin{split}
    &||X^{t+1}(\delta) \bm{s}^{t+1}(\delta)- \mu^{t+1}(\delta) 1|| =\\
    & \left|\left| (X^t \bm{s}^t -\mu^t\bm{1}) + \delta \left( -X^t \bm{s}^t  + \mu^t\bm{1} \right)\right.\right.\\ &+\left.\left.\delta^2\left(D_x \bm{d_s} + \frac{\bm{d_x}^T \bm{\epsilon_x} + \bm{d_s}^T \bm{\epsilon_x}}{n+1}\right)\right|\right|\leq \\
    &||(1-\delta)(X^t \bm{s}^t -\mu^t \bm{1})|| + \delta^2 ||D_x \bm{d_s}|| + \delta^2 O(\epsilon).
\end{split}
\end{equation}
 For the next steps we follow the procedure indicated in lemma 4 and theorem 1 of \cite{Predictor-CorrectorI}. According to them, one chooses $\delta$ such that 
 \begin{equation}
 \delta^2||D_x \bm{d_s}||\leq \mu^t/8. \label{choosing delta}
 \end{equation}
Recall that in the last line of \eqref{norm of Xs - mu delta}, by hypothesis the first term is smaller than $(1-\delta)\mu^t/4$, because the previous step was a corrector step. With the $\delta$ from \eqref{choosing delta}, \eqref{beta neighbourhood} is fulfilled, as proved in Lemma 4 from \cite{Predictor-CorrectorI}. Then we must calculate the size of $||D_x \bm{d_s}||$. Using the same methods as theorem 1 of \cite{Predictor-CorrectorI}, we get
 \begin{equation}
     ||D_x \bm{d_s}||\leq \frac{\sqrt{2}}{4} ((n+1)\mu^{t+1} + ||\bm{\epsilon}||^2),
 \end{equation}
for $\bm{\epsilon}$ a vector that also has entries of size $O(\epsilon)$. Thus $||\bm{\epsilon}||^2 = (n+1) O(\epsilon)$. Using \eqref{choosing delta} we deduce that $\delta \geq 8^{-1/4}(n+1)^{-1/2}\sqrt{\mu^t/(\mu^t + O(\epsilon))}$. The fact that $\delta \geq O(n^{-1/2})$ is key to obtain a number of steps that grows as $O(n^{1/2})$ as explained in theorem 1 of \cite{Predictor-CorrectorI} and theorem 6 of \cite{Predictor-Corrector}. Consequently, we can see that the step size in the Predictor step gets almost the same convergence guarantees as in the exact arithmetic case. Notice that the only caveat is that we will not be able to reduce the duality gap more than $\mu = O(\epsilon)$, but that is fine. This concludes our algorithm.

\pagebreak
\begin{widetext}
\section{\label{sec:Structure of the algorithm}Overall structure of the algorithm}

\subsection{Initialization}
The initialization procedure consists in preparing the matrix $M$, and the state $f$. 
\begin{algorithm}[H]
\caption{Quantum interior point algorithm initialization.}\label{QIPInit}
\begin{algorithmic}[1]
\Procedure{Initialization}{}

\BState\textbf{Problem:} Solve the following dual problems
\begin{equation}
\text{minimize } c^T x, \qquad \text{subject to } A x \geq b, \qquad x\geq 0. 
\end{equation}
and
\begin{equation}
\text{maximize } b^T y,\qquad \text{subject to } A^T y \leq c. 
\end{equation}

\BState\textbf{Input:} Sparse matrix $A\in\mathbb{R}^{m\times n}$, sparse vector $c\in\mathbb{R}^{m}$, vector $b\in\mathbb{R}^{n}$.
\BState\textbf{Output:} Dual solutions $y\in\mathbb{R}^{m}$ and $x\in\mathbb{R}^{n}$, or a signal that the problem is infeasible.
\BState\textbf{Initialization:} Want to form the matrix \eqref{Matrix system}.
\State Define $\tau=k=\theta=1$.
\State Set $\bm{x}^0= \bm{s}^0= 1_{n\times 1}$, and $\bm{y}^0=0_{m\times 1}$.
\State Calculate $\bar{z}$.
\State Calculate $\bar{\bm{b}}$ and $\bar{\bm{c}}$.
\State Set $t=0$.
\State Create the quantum-accessible classical data structure for $M^0$.
\EndProcedure
\end{algorithmic}
\end{algorithm}

\subsection{Termination}
In the termination we propose one possible way of using Grover to run the termination explained in \cite{Predictor-Corrector}. Any other classical termination is also possible.

\begin{algorithm}[H]
\caption{Quantum interior point algorithm termination.}\label{QIPTerm}
\begin{algorithmic}[1]
\Procedure{Termination}{}
\BState In this section we propose a termination technique using Grover algorithm \cite{Grover} and \cite{GroverMin} to find the optimal solution. We suppose the search space is small enough to allow for this `brute force' search without affecting the complexity class of the main loop. This technique can be nevertheless substituted by any other efficient classical termination.
\If {termination of algorithm \ref{QIPLoop} was due to $2^{nd}$ criterion}
\State \eqref{LP} or \eqref{LD} do not have feasible solutions such that $||(\bm{x},^T\bm{s}^T)||\leq 1/(2\epsilon_3)-1$. The problem is infeasible or unbounded. Check feasibility with the latest available step.
\EndIf
\If {termination of algorithm \ref{QIPLoop} was due to $1^{st}$ criterion}
\If{$\tau^t\geq k^t$}
\State Use Grover search algorithm \cite{Grover} to find all possible solutions to \eqref{termination 1b}, without reading them out. 
\State Use Grover Search minimum finding algorithm \cite{GroverMin} to find the minimum of the possible states.
\EndIf
\If{$\tau^t< k^t$}
\State Use Grover search algorithm \cite{Grover} to find all possible solutions to \eqref{termination 2b}, without reading them out. 
\State Use Grover Search minimum finding algorithm \cite{GroverMin} to find the minimum of the possible states.
\EndIf
\EndIf
\EndProcedure
\end{algorithmic}
\end{algorithm}

\pagebreak

\subsection{Main loop}
The main loop consists in two steps called predictor and corrector. The structure of them is very similar: 
\begin{enumerate}
    \item Update the data structures for $f^t$ and $M^t$.
    \item Prepare $\ket{f}$ and solve $M\ket{d}=\ket{f}$ with QLSA.
    \item Read $\ket{d}\rightarrow d$ and calculate the new vector $\bm{v}^{t+1}=(\bm{y}^{t+1},\bm{x}^{t+1},\tau^{t+1},\theta^{t+1},\bm{s}^{t+1},k^{t+1})$
\end{enumerate}

\begin{algorithm}[H]
\caption{Quantum interior point algorithm loop.}\label{QIPLoop}
\begin{algorithmic}[1]
\Procedure{Main Loop}{}
\BState\textbf{Main loop:} Loop $O(L\sqrt{n})$ times over $t$ until one of the following two criteria are fulfilled: Choose $\epsilon_1,\epsilon_2,\epsilon_3$ small numbers and 
\begin{enumerate}
    \item $(\bm{x}^t/\tau^t)^T( \bm{s}^t/\tau^t)\leq \epsilon_1$ and $(\theta^t/\tau^t)||(\bar{\bm{b}}^T,\bar{\bm{c}}^T)||\leq \epsilon_2$.
    \item $\tau^t \leq \epsilon_3$.
\end{enumerate}
We will have to iterate $O(\bar{t})$ times: $\bar{t}=\max[\log((\bm{x}^0)^T( \bm{s}^0)/(\epsilon_1\epsilon_3^2)), \log(||(\bar{\bm{b}}^T,\bar{\bm{c}}^T)||/\epsilon_2\epsilon_3)]$.

\BState{}\textit{Update of the data structures:}
\State Update the data structures that save $M^t$ and $f^t$  with $\gamma^t=0$.

\BState \textit{Predictor step:}
\State Use theorem \ref{QLSA theorem} as a QLSA to solve \eqref{Matrix system} $O((n+m)\epsilon^{-2})$ times.
\State Read the results using the tomography algorithm \ref{Tomograpy}.
\State Check the classical answer using Swap Test with the correct answer. If not, iterate the two previous steps.
\State Rescale the vector using the norm of the solution obtained using theorem \ref{QLSA theorem}.
\State Correct the global sign multiplying a row of $A$ with the solution and comparing it against one of the entries of $\bm{f}^t$.
\State For each coordinate $i$ shift ${d_x}_i$ (respectively ${d_s}_i$) $\epsilon'$ when $x_i> s_i$ ($x_i> s_i$) so that \eqref{eq12} is fulfilled exactly.
\State Use binary search to find the $\delta$ that fulfills that \eqref{predictor sum} $\in \mathcal{N}(1/2)$.
\State Calculate the values of $(\bm{y}^{t+1},\bm{x}^{t+1},\tau^{t+1},\theta^{t+1},\bm{s}^{t+1},k^{t+1})$ using
\eqref{predictor sum}.
\State $t \leftarrow t+1$.

\BState{}\textit{Update of the data structures:}
\State Update the data structures that save $M^t$ and $f^t$  with $\gamma^t=1$.

\BState \textit{Corrector step:}

\State Use theorem \ref{QLSA theorem} as a QLSA to solve \eqref{Matrix system} $O((n+m)\epsilon^{-2})$ times.
\State Read the results using the tomography algorithm \ref{Tomograpy}.
\State Check the classical answer using Swap Test with the correct answer. If not, iterate the two previous steps.
\State Rescale the vector using the norm of the solution obtained using theorem \ref{QLSA theorem}.
\State Correct the global sign multiplying a row of $A$ with the solution and comparing it against one of the entries of $\bm{f}^t$.
\State Calculate the values of $(\bm{y}^{t+1},\bm{x}^{t+1},\tau^{t+1},\theta^{t+1},\bm{s}^{t+1},k^{t+1})$ using
\eqref{corrector sum}.
\State If the new point $(\bm{y}^{t+1},\bm{x}^{t+1},\tau^{t+1},\theta^{t+1},\bm{s}^{t+1},k^{t+1})\notin \mathcal{N}(1/4)$ use gradient descent with parameter $\epsilon'' = O(\epsilon')$ determined by \eqref{calculation of epsilon''} to shift it slightly until it is inside $\mathcal{N}(1/4)$.
\State $t \leftarrow t+1$.
\EndProcedure
\end{algorithmic}

\end{algorithm}
\end{widetext}

\end{document}